\documentclass{aa}

\usepackage{graphicx}
\usepackage{epsfig}

\newcommand{\pl}{\partial}
\renewcommand{\d}{{\rm d}}
\newcommand{\inta}{\int_{-i\infty}^{+i\infty}}
\newcommand{\beq}{\begin{equation}}
\newcommand{\eeq}{\end{equation}}
\newcommand{\beqa}{\begin{eqnarray}}
\newcommand{\eeqa}{\end{eqnarray}}
\newcommand{\bea}{\begin{array}}
\newcommand{\ea}{\end{array}}
\newcommand{\bx}{{\bf x}}
\newcommand{\br}{{\bf r}}
\newcommand{\bv}{{\bf v}}

\newcommand{\cG}{{\cal G}}

\newcommand{\rhob}{\overline{\rho}}
\newcommand{\bk}{{\bf k}}
\newcommand{\lag}{\langle}
\newcommand{\rag}{\rangle}
\newcommand{\Om}{\Omega_{\rm m}}
\newcommand{\Ol}{\Omega_{\Lambda}}

\newcommand{\dR}{\delta_{R}}
\newcommand{\dL}{\delta_L}
\newcommand{\DL}{\Delta_L}

\newcommand{\cF}{{\cal F}}
\newcommand{\xib}{\overline{\xi}}
\newcommand{\Det}{{\rm Det}}

\newcommand{\cD}{{\cal D}}
\newcommand{\psib}{\overline{\psi}}

\newcommand{\cFr}{{\cal F}_{\rho}}
\newcommand{\cGr}{{\cal G}_{\rho}}

\newcommand{\om}{\omega}
\newcommand{\gam}{\gamma}
\newcommand{\psit}{\tilde{\psi}}
\newcommand{\phit}{\tilde{\varphi}}

\newcommand{\at}{\tilde{a}}
\newcommand{\omt}{\tilde{\omega}}

\newcommand{\Pt}{\tilde{\cal P}}
\newcommand{\rhot}{\tilde{\rho}}
\newcommand{\yt}{\tilde{y}}
\newcommand{\vp}{v_{\perp}}
\newcommand{\bL}{{\bf L}}
\newcommand{\Omr}{\Omega_r}
\newcommand{\Omt}{\Omega_{\theta}}
\newcommand{\Omo}{\Omega_0}
\newcommand{\cE}{{\cal E}}
\newcommand{\dc}{\delta_c}
\newcommand{\Dc}{\Delta_c}
\newcommand{\Nb}{\overline{N}}
\newcommand{\cP}{{\cal P}}

\begin{document}

 

\title{Dynamics of gravitational clustering IV. The probability distribution of rare events.}   
\author{P. Valageas}  
\institute{Service de Physique Th\'eorique, CEN Saclay, 91191 Gif-sur-Yvette, France} 
\date{Received / Accepted }

\abstract{
Using a non-perturbative method developed in a previous article (\cite{paper2}) we investigate the tails of the probability distribution $\cP(\rho_R)$ of the overdensity within spherical cells. Since our approach is based on a steepest-descent approximation which should yield exact results in the limit of rare events it applies to all values of the rms linear density fluctuation $\sigma$, from the quasi-linear up to the highly non-linear regime. First, we derive the low-density tail of the pdf. We show that it agrees with perturbative results when the latter are finite up to the first subleading term, that is for linear power-spectra $P(k) \propto k^n$ with $-3<n<-1$. Over the range $-1<n<1$ some shell-crossing occurs (which leads to the break-up of perturbative approaches) but this does not invalidate our approach. In particular, we explain that we can still obtain an approximation for the low-density tail of the pdf. This feature also clearly shows that perturbative results should be viewed with caution even when they are finite. We point out that our results can be recovered by a simple spherical model (this is related to the spherical symmetry of our problem). On the other hand, we show that this low-density tail cannot be derived from the stable-clustering ansatz in the regime $\sigma \gg 1$ since it involves underdense regions which are still expanding. Second, turning to high-density regions we explain that a naive study of the radial spherical dynamics fails. Indeed, a violent radial-orbit instability leads to a fast relaxation of collapsed halos (over one dynamical time) towards a roughly isotropic equilibrium velocity distribution. Then, the transverse velocity dispersion stabilizes the density profile so that almost spherical halos obey the stable-clustering ansatz for $-3<n<1$. We again find that our results for the high-density tail of the pdf agree with a simple spherical model (which takes into account virialization). Moreover, they are consistent with the stable-clustering ansatz in the non-linear regime. Besides, our approach justifies the large-mass cutoff of the Press-Schechter mass function (although the various normalization parameters should be modified). Finally, we note that for $\sigma \ga 1$ an intermediate region of moderate density fluctuations appears which calls for new non-perturbative tools.
}

\maketitle

\section{Introduction}

In usual cosmological scenarios large-scale structures in the universe have formed through the growth of small initial density fluctuations by gravitational instability (e.g., \cite{Peeb1}). Moreover, in most cases of cosmological interest the amplitude of these density fluctuations increases at smaller scales, as in the standard CDM model. This leads to a hierarchical scenario of structure formation where smaller scales enter the non-linear regime first, building small virialized objects which later become part of increasingly large and massive structures. These halos will produce galaxies or clusters of galaxies (depending on non-gravitational processes like collisional cooling) which build a complex network among large voids. Thus, in order to describe the non-linear structures we observe in the present universe it is of great interest to understand the non-linear evolution of the density field.

Unfortunately, this is a rather difficult task since this non-linear regime (i.e. small scales or rare large density fluctuations on large scales) cannot be described by perturbative methods (except large voids for $n<-1$). In fact, very few results have been obtained so far in this domain since most rigorous approaches to the dynamics of gravitational clustering have relied on perturbative methods. Therefore, they were restricted to the early stages of the non-linear evolution. In this paper, we make use of a non-perturbative method developed in a previous paper (\cite{paper2}) to tackle this fully non-linear regime. More precisely, since this approach is based on a steepest-descent approximation we investigate the regime of rare events where it should yield asymptotically exact results. Thus, we consider the very high density and low density tails of the probability distribution function (pdf) $\cP(\rho_R)$ of the overdensity within spherical cells. Then, this approach applies to any value of the rms linear density fluctuation $\sigma$ so that it describes all regimes, from linear to highly non-linear scales. However, while in the quasi-linear regime it is able to predict the whole pdf (this was investigated in a specific manner in \cite{paper2}) in the highly non-linear regime it only provides the rare-event tails of the pdf.

This article is organized as follows. First, in Sect.\ref{Action} we recall the path-integral formalism developed in \cite{paper2} which provides an explicit expression for the pdf $\cP(\rho_R)$ in terms of the initial conditions. Then, in Sect.\ref{Rare underdensities} we investigate the very low density tail of the pdf. We first derive the saddle-point of the action which governs this regime in Sect.\ref{Spherical saddle-point} and we give in Sect.\ref{Low-density tail of the pdf} the low density tail of the pdf this implies. We compare in Sect.\ref{Comparison with previous works} our results with perturbative methods, a simple spherical model and the usual stable-clustering ansatz. Next, in Sect.\ref{Rare overdensities} we turn to the high-density tail of the pdf. We first derive in Sect.\ref{Spherical collapse} and Sect.\ref{Spherical overdense saddle-point} a naive spherical saddle-point but we show in Sect.\ref{Virialization processes} that virialization processes are sufficiently violent to significantly modify this simple approach. Then, we give in Sect.\ref{High density tail of the pdf} the high-density tail of the pdf obtained by a more careful study. We finally compare this result with the stable-clustering ansatz in Sect.\ref{Comparison with the non-linear scaling model} and with the standard Press-Schechter mass function (\cite{PS}) in Sect.\ref{Press-Schechter prescription}.

\section{Action $S[\dL]$}
\label{Action}

In this section, we introduce our notations and we briefly recall the formalism developed in \cite{paper2} which allows us to evaluate the pdf $\cP(\dR)$. More precisely, we investigate the statistical properties of the overdensity $\rho_R$ within spherical cells of comoving radius $R$, volume $V$:
\beq
\rho_R \equiv 1 + \dR \;\;\; \mbox{with} \;\;\; \dR \equiv \int_V \frac{\d^3x}{V} \; \delta(\bx) .
\label{rhoR}
\eeq
Here $\delta(\bx)$ is the non-linear density contrast at the comoving coordinate $\bx$, at the time of interest. We investigate in this article the case of Gaussian initial conditions, so that the statistical properties of the random linear density field $\dL(\bx)$ are fully defined by the two-point correlation:
\beq
\DL(\bx_1,\bx_2) \equiv \lag \dL(\bx_1) \dL(\bx_2) \rag .
\label{Dl1}
\eeq
The kernel $\DL$ is symmetric, homogeneous and isotropic: $\DL(\bx_1,\bx_2) = \DL(|\bx_1-\bx_2|)$. Moreover, the matrix $\DL$ can be inverted and its inverse $\DL^{-1}$ is positive definite since we have:
\beq
\dL . \DL^{-1} . \dL = \int \d\bk \; \frac{|\dL(\bk)|^2}{P(k)}
\label{Dl2}
\eeq
for real fields $\dL(\bx)$, see \cite{paper2}. Here we introduced the short-hand notation:
\beq
f_1 . \DL^{-1} . f_2 \equiv \int \d\bx_1 \d\bx_2 \; f_1(\bx_1) . \DL^{-1}(\bx_1,\bx_2) . f_2(\bx_2)
\label{Dl0}
\eeq
for any real fields $f_1$ and $f_2$. Moreover, the power-spectrum $P(k)$ of the linear density contrast used in eq.(\ref{Dl2}) is given by:
\beq
\lag \dL(\bk_1) \dL(\bk_2) \rag \equiv P(k_1) \; \delta_D(\bk_1+\bk_2)
\label{Pk1}
\eeq
where we defined the Fourier transform of the linear density field by:
\beq
\dL(\bx) = \int \d\bk \; e^{i \bk.\bx} \; \dL(\bk) .
\label{Four1}
\eeq
Finally, it is convenient to introduce the usual rms linear density fluctuation $\sigma(R)$ in a cell of radius $R$:
\beq
\sigma^2(R) \equiv \lag \delta_{L,R}^2 \rag = \int_V \frac{\d\bx_1}{V} \frac{\d\bx_2}{V} \; \DL(\bx_1,\bx_2)
\label{sig1}
\eeq
where $\delta_{L,R}$ is the mean linear density contrast over the volume $V$ as in eq.(\ref{rhoR}).

The previous expressions describe the statistical properties of the initial conditions, through the linearly-evolved density field $\dL(\bx)$ (see also \cite{paper1}). Then, if we could explicitly solve the collisionless Boltzmann equation which governs the gravitational dynamics (coupled to the Poisson equation) this would provide the statistics of the actual non-linear density field $\delta(\bx)$. In any case, we can always write a closed formal expression for the pdf $\cP(\rho_R)$ of the overdensity $\rho_R$, following the method presented in \cite{paper2}. To do so, it is convenient to introduce the Laplace transform $\psi(y)$ of the pdf, given by:
\beq
\psi(y) \equiv \lag e^{-y \rho_R} \rag \equiv \int_{0}^{\infty} \d\rho_R \; e^{-y \rho_R} \; \cP(\rho_R)
\label{psi1}
\eeq
since $\rho_R \geq 0$. Here, the symbol $\lag .. \rag$ expresses the average over the initial conditions. Then, the pdf can be recovered from $\psi(y)$ through the inverse Laplace transform:
\beq
\cP(\rho_R) = \inta \frac{\d y}{2\pi i} \; e^{y \rho_R} \; \psi(y) .
\label{P1}
\eeq
As described in \cite{paper2}, we can write an explicit expression for the average over the initial conditions which appears in eq.(\ref{psi1}). Indeed, since these initial conditions can be fully described by the linear growing mode $\dL(\bx)$ (see \cite{paper1}) which is a Gaussian random field we can write:
\beq
\psi(y) = \left( \Det\DL^{-1} \right)^{1/2} \int [\d\dL(\bx)] \; e^{-y \rho_R[\dL] -\frac{1}{2} \dL . \DL^{-1} . \dL} 
\label{psi2}
\eeq
where the inverse kernel $\DL^{-1}$ was introduced in eq.(\ref{Dl2}). Here the functional $\rho_R[\dL]$ is the non-linear overdensity $\rho_R$ produced by the linear density field $\dL$.

As in \cite{paper2}, it is actually convenient to introduce the rescaled generating function $\psib(y)$ defined by:
\beq
\psi(y) \equiv \psib \left( y \; \sigma^2(R) \right) .
\label{psib1}
\eeq
Indeed, this allows us to factorize the amplitude of the power-spectrum $P(k)$ in the ``action'' $S[\dL]$ which characterizes our system. Thus, we can now write eq.(\ref{psi2}) as:
\beq
\psib(y) = \left( \Det\DL^{-1} \right)^{1/2} \int [\d\dL(\bx)] \; e^{- S[\dL] /\sigma^2(R)}
\label{psib2}
\eeq
where we introduced the action $S[\dL]$:
\beq
S[\dL] = y \; \rho_R[\dL] + \frac{\sigma^2(R)}{2} \; \dL . \DL^{-1} . \dL
\label{S1}
\eeq
The action $S[\dL]$ is independent of the normalization of the linear power-spectrum $P(k)$ since $\DL \propto \sigma^2$, see eq.(\ref{sig1}). In \cite{paper2} the change of variable $y \rightarrow y/\sigma^2$ in eq.(\ref{psib1}) was crucial since it allowed us to show that the steepest-descent method was asymptotically exact in the limit $\sigma \rightarrow 0$. Indeed, it is clear that in this quasi-linear limit the path-integral in eq.(\ref{psib2}) is given by the global minimum of the action $S$. However, in this article we no longer study this quasi-linear limit. Indeed, we investigate the regime of rare events, i.e. the limits $\rho_R \rightarrow 0$ (rare ``voids'') or $\rho_R \rightarrow +\infty$ (very high overdensities), and we take $\sigma$ to be finite. Thus, our study applies both to the linear and non-linear regimes since the value of $\sigma$ is actually irrelevant. Then, the change of variable introduced in eq.(\ref{psib1}) is no longer essential and we could directly work with $\psi(y)$. Nevertheless, it is still convenient to work with the rescaled generating function $\psib(y)$ since it allows us to get rid of the amplitude of the rms density fluctuations which plays no key role in the physics we investigate here. Besides, it allows us to use the same expressions as in \cite{paper2}.

\section{Rare underdensities}
\label{Rare underdensities}

We first consider the statistical properties of very rare underdensities (i.e. ``voids''), using the tools laid out in the previous section. Indeed, this is much easier than the study of large overdensities which involves shell-crossing in an essential way. Hence it is convenient to start with voids to recall the basic ideas behind the steepest-descent method introduced in \cite{paper2} since it only requires minor modifications to be applied to underdensities. We shall investigate the high-density tail of the pdf $\cP(\rho_R)$ in a next section.

In the following we restrict ourselves to the case of a critical-density universe. Then, the spherical solution of the collisionless Boltzmann equation is explicitly known, as long as there is no shell-crossing.

\subsection{Spherical saddle-point}
\label{Spherical saddle-point}

\begin{figure}
\centerline{\epsfxsize=8cm \epsfysize=5.8cm \epsfbox{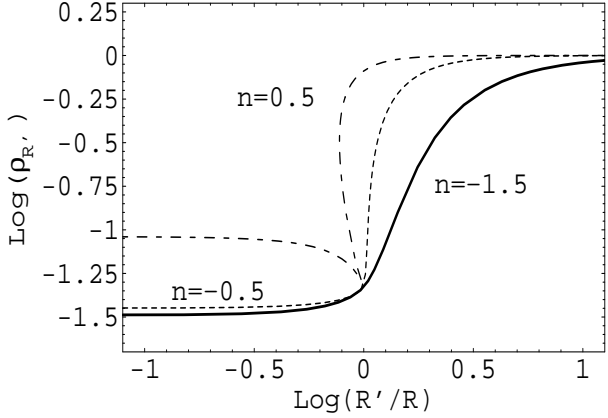}}
\caption{Cumulative non-linear overdensity profile $\rho_{R'}$ of the spherical saddle-point. The solid line corresponds to $n=-1.5$, the dashed-line to $n=-0.5$ and the dot-dashed line to $n=0.5$. For the curve $n=0.5$ we clearly see that shell-crossing occurs at $R' \sim R$.} 
\label{figprofil}
\end{figure}

Thus, our goal is now to evaluate the path-integral (\ref{psib2}) in the limit $\rho_R \rightarrow 0$ in order to derive the low-density tail of the pdf $\cP(\rho_R)$. To this order, we can try a steepest-descent approximation, in the spirit of the calculation performed in \cite{paper2}. Since $\rho_R[\dL] = 1+\dR[\dL]$ the action $S[\dL]$ is directly related to the action studied in \cite{paper2} and we can use the results obtained in that previous work. In particular, if there is no shell-crossing we know that the action $S[\dL]$ admits a spherically symmetric saddle-point $\dL(\bx)$ given by:
\beq
\delta_{L,R_L} = - y \; \frac{ \cFr' \left[ \delta_{L,R_L} \right] \sigma^2(R_L)/\sigma^2(R) }{ 1 - \cFr' \left[ \delta_{L,R_L} \right] \frac{R^3}{3R_L^2} \delta_{L,R_L} \frac{1}{\sigma(R_L)} \; \frac{\d \sigma}{\d R}(R_L) } 
\label{col1}
\eeq
together with:
\beq
\dL(\bx) =  \delta_{L,R_L} \int_{V_L} \frac{\d\bx'}{V_L} \; \frac{\DL(\bx,\bx')}{\sigma^2(R_L)} .
\label{col2}
\eeq
Here the Lagrangian comoving radius $R_L$ is such that the matter enclosed within this volume $V_L$ in the primordial universe (i.e. $M_L = 4\pi/3 \; \rhob R_L^3$) ends up within the radius $R$ in the actual non-linear density field. Moreover, for such spherically symmetric initial conditions the actual non-linear overdensity $\rho_R$ only depends on the linear density contrast $\delta_{L,R_L}$ within the radius $R_L$ through the function $\cFr$. This function is the usual spherical solution of the dynamics. It can be expressed in terms of hyperbolic functions as (for $\Om=1$, see \cite{Peeb1}):
\beq 
\left\{ \begin{array}{l}
\cFr(\dL) = {\displaystyle \frac{9}{2} \; \frac{(\sinh
\eta-\eta)^2}{(\cosh \eta-1)^3} } \\ \\ \delta_L = {\displaystyle -
\frac{3}{20} \; \left[ 6 (\sinh \eta-\eta) \right]^{2/3} } \end{array}
\right.
\label{Fcol1}
\eeq
Note that since we study $\rho_R=1+\dR$ rather than $\dR$ the function $\cFr$ defined in eq.(\ref{Fcol1}) differs from the usual function $\cF$ (used in \cite{paper2}) by a factor $+1$. Besides, the non-linear quantities $\rho_R$ and $R$ are related to the linear variables $\delta_{L,R_L}$ and $R_L$ by:
\beq
\left\{ \begin{array}{l}
{\displaystyle \rho_R = \cFr \left[ \delta_{L,R_L} \right] } \\ \\
{\displaystyle R_L^3 = \rho_R \; R^3 }
\end{array} \right.
\label{F1}
\eeq
Then, the implicit eq.(\ref{col1}) defines the normalization $\delta_{L,R_L}$ of the spherical saddle-point associated to a given value of $y$, while eq.(\ref{col2}) provides the density profile of this saddle-point. The cumulative linear density profile $\delta_{L,R_L'} / \delta_{L,R_L}$ of this spherical saddle-point is displayed in Fig.1 in \cite{paper2}. We shall simply recall here that for a power-law linear power-spectrum $P(k) \propto k^n$ we have the asymptotic behaviours:
\beq
R_L' \rightarrow 0 : \;\; \frac{\delta_{L,R_L'}}{\delta_{L,R_L}} \rightarrow 2^n \; \frac{(1-n)(3-n)}{3}
\label{profil3}
\eeq
and
\beq
R_L' \rightarrow \infty : \; \frac{\delta_{L,R_L'}}{\delta_{L,R_L}} \sim 2^n \; \frac{(1-n)(3-n)}{3}  \left( \frac{R_L}{R_L'} \right)^{n+3} .
\label{profil4}
\eeq
We display in Fig.\ref{figprofil} the cumulative non-linear overdensity profile $\rho_{R'}$. It is obtained from the relation $\rho_{R'} = \cFr [ \delta_{L,R_L'} ]$ which applies to all radii $(R',R_L')$, where $R_L'$ is the linear Lagrangian radius associated with the actual non-linear radius $R'$ (assuming there is no shell-crossing).

Of course, as for the linear density contrast $\delta_{L,R_L'}$ we find that $\rho_{R'}$ remains finite for small $R'$, of the order of $\rho_R$, while it goes to unity (i.e. the density contrast vanishes) at large radius $R'$. Besides, since we now probe the highly non-linear regime we must pay attention to shell-crossing. Here, we must recall that the previous results were derived with the assumption that there is no shell-crossing, so that the simple mapping (\ref{F1}) is valid. However, it is clear in Fig.\ref{figprofil} that the profile obtained for the case $n=0.5$ leads to shell-crossing at $R' \sim R$. The condition which expresses that shell-crossing appears is $\d R'/\d R_L'=0$. Using this constraint in the system (\ref{F1}) written for an arbitrary pair $(R',R_L')$ we obtain:
\beq
\mbox{shell-crossing :} \;\; \left| \frac{\d \ln \cFr}{\d \ln (-\delta_{L,R_L'})} \right| \geq 3 \;  \left| \frac{\d \ln R_L'}{\d \ln (-\delta_{L,R_L'})} \right| .
\label{cros1}
\eeq
Thus, we need the behaviour of the function $\cFr(\dL)$ for large negative $\dL$. From eq.(\ref{Fcol1}) we obtain at large $\eta$:
\beq
\dL \rightarrow -\infty : \;\; \cFr(\dL) \simeq \left( - \frac{20}{27} \; \dL \right)^{-3/2} .
\label{asymp1}
\eeq
Substituting this result into eq.(\ref{cros1}) we get the condition:
\beq
\mbox{shell-crossing :} \;\; \frac{\d \ln (-\delta_{L,R_L'})}{\d \ln R_L'} \leq -2 .
\label{cros2}
\eeq
Then, from the asymptotic behaviour (\ref{profil4}) we find that for $n>-1$ some shell-crossing occurs at scales $R' \ga R$ for very underdense saddle-points. In Fig.\ref{figprofil} the curve shown for the case $n=-0.5$ does not exhibit any shell-crossing yet because the normalization $\delta_{L,R_L}$ is not large enough but we can check numerically that for a more negative value of $\delta_{L,R_L}$ ($\rho_R \la 10^{-2}$) some shell-crossing indeed appears. For $n \geq 0$ some shell-crossing also occurs at $R' \sim R$ since for such power-spectra the local density contrast $\dL(\bx)$ changes sign at the radius $R_L$. Then the derivative of the cumulative density contrast $\delta_{L,R_L'}$ is discontinuous at $R_L$ and $\d \ln (-\delta_{L,R_L'}) / \d \ln R_L' \leq -3$ for $R_L' > R_L$.

These results imply that the spherical saddle-point obtained in eq.(\ref{col1}) and eq.(\ref{col2}) is not correct for large negative $\delta_{L,R_L}$ if $n>-1$. However, since shell-crossing only appears over a limited range of radii of the order of $R_L$ we can expect the previous state $\dL(\bx)$ to be a reasonable approximation to the exact spherically symmetric saddle-point. Moreover, for $-1<n<0$ shell-crossing only occurs for $R_L'>R_L$ so that the radial profile (\ref{col2}) is correct for $R_L'<R_L$ and it should only be modified at $R_L'>R_L$. In order to derive the exact saddle-point we need to explicitly take into account shell-crossing which makes the calculation much more difficult, since we do not know the exact functional $\rho_R[\dL]$ in this regime. However, because of the spherical symmetry of the problem which we consider here, the saddle-point should remain spherically symmetric. In the following, we shall use the linear state $\dL(\bx)$ described by eq.(\ref{col1}) and eq.(\ref{col2}) even for $n>-1$. In this latter case, our results are no longer exact but we can expect that they should still provide a good approximation to the actual pdf.

\subsection{Generating function}
\label{Generating function}

Using the spherically symmetric saddle-point obtained in Sect.\ref{Spherical saddle-point} we can use a steepest-descent approximation to the path-integral (\ref{psib2}). That is, we approximate this path-integral by the Gaussian integration around the spherical saddle-point $\dL(\bx)$. This yields:
\beq
\psib(y) \simeq \left( \Det\DL^{-1} \right)^{1/2} \; \left( \Det M_y \right)^{-1/2} \; e^{- S_y /\sigma^2(R)}
\label{psib3}
\eeq
where the matrix $M_y$ is the Hessian of the exponent evaluated at the saddle-point $\dL$ obtained in Sect.\ref{Spherical saddle-point}:
\beqa
\lefteqn{ M_y(\bx_1,\bx_2) \equiv \frac{\delta^2 (S/\sigma^2)}{\delta(\dL(\bx_1)) \delta(\dL(\bx_2))} } \nonumber \\ & & = \frac{y}{\sigma^2(R)} \; \frac{\delta^2 (\dR)}{\delta(\dL(\bx_1)) \delta(\dL(\bx_2))} +  \DL^{-1}(\bx_1,\bx_2) . 
\label{M1}
\eeqa
In eq.(\ref{M1}) we used the fact that the second derivative of $\dR$ is also the second derivative of $\rho_R$ since $\rho_R=1+\dR$. The quantity $S_y$ which appears in eq.(\ref{psib3}) is the value of the action $S[\dL]$ defined in eq.(\ref{S1}) at the spherical saddle-point $\dL(\bx)$ derived in Sect.\ref{Spherical saddle-point}. As in \cite{paper2}, after we substitute eq.(\ref{col2}) into eq.(\ref{S1}) we can write $S_y$ as the solution of the implicit system:
\beq
\left\{ \begin{array}{l}
{\displaystyle \tau = - y \; \cGr'(\tau) } \\ \\
{\displaystyle S_y = y \; \cGr(\tau) + \frac{\tau^2}{2} }
\end{array} \right.
\label{Sy1}
\eeq
where we introduced the variable $\tau$ and the function $\cGr(\tau)$ defined by:
\beq
\tau(\delta_{L,R_L}) \equiv \frac{- \; \delta_{L,R_L} \; \sigma(R)}{\sigma \left[ \cFr(\delta_{L,R_L})^{1/3} R \right] }
\label{tau1}
\eeq
and:
\beq
\cGr(\tau) \equiv \cFr\left(\delta_{L,R_L}\right) = \rho_R .
\label{G1}
\eeq
Using eq.(\ref{tau1}) we see that the function $\cGr(\tau)$ defined in eq.(\ref{G1}) also obeys the implicit equation:
\beq
\cGr(\tau) = \cFr\left[ - \tau \; \frac{\sigma\left[\cGr(\tau)^{1/3}R\right]}{\sigma(R)} \right] .
\label{G2}
\eeq
As for $\cFr$, the function $\cGr$ differs from the one introduced in \cite{paper2} by a factor $+1$ because we study here $\rho_R$ rather than $\dR$. For a power-law power-spectrum $P(k) \propto k^n$ we have $\sigma(R) \propto R^{-(n+3)/2}$ so that eq.(\ref{G2}) simplifies to:
\beq
\cGr(\tau) = \cFr\left[ - \tau \; \cGr(\tau)^{-(n+3)/6} \right] .
\label{G3}
\eeq

The previous equations allow us to evaluate the generating function $\psib(y)$ through eq.(\ref{psib3}). However, as for the quasi-linear regime studied in \cite{paper2} we must first ensure that the spherical saddle-point derived in Sect.\ref{Spherical saddle-point} is indeed the global minimum of the action $S[\dL]$. Since we investigate here another regime ($\rho_R \rightarrow 0$ and not $\sigma \rightarrow 0$) we have to consider this point anew. Besides, we need to make sure that the Gaussian approximation really makes sense. Indeed, contrary to what occurs in the quasi-linear regime, the action $S[\dL]$ is not multiplied by a large factor in the exponent in eq.(\ref{psib2}) (we do not study the limit $1/\sigma^2 \rightarrow \infty$) so that it is not obvious a priori that the path-integral should be dominated by a small neighbourhood of the saddle-point $\dL$ in the limit $\rho_R \rightarrow 0$.

\subsection{Validity of the steepest-descent method}
\label{Validity of the steepest-descent method}

First, we note that the regime of rare voids we investigate here corresponds to large positive $y$. This is obvious from the definition (\ref{psi1}) which clearly shows that for $y \rightarrow +\infty$ we mainly probe the behaviour of the pdf $\cP(\rho_R)$ for small overdensities $\rho_R \rightarrow 0$, that is for very underdense regions. Of course, rare voids also correspond to $\delta_{L,R_L} \rightarrow -\infty$, hence to $\tau \rightarrow +\infty$ as can be seen from eq.(\ref{tau1}). Moreover, the behaviour of $\cFr(\delta_{L,R_L})$ in this regime was derived in eq.(\ref{asymp1}). Using eq.(\ref{G3}) this yields:
\beq
\tau \rightarrow +\infty : \;\; \rho_R = \cGr(\tau) \simeq \left( \frac{20}{27} \; \tau \right)^{-6/(1-n)} .
\label{asytau1}
\eeq
Then, using eq.(\ref{Sy1}) we obtain:
\beq
y \rightarrow +\infty : \;\; \tau \simeq \left( \frac{20}{27} \right)^{-3/(4-n)} \left( \frac{6 \;y}{1-n} \right)^{(1-n)/(8-2n)} .
\label{tauy1}
\eeq

As explained above, in order to justify the steepest-descent method we first need to show that the spherical saddle-point derived in Sect.\ref{Spherical saddle-point} is the global minimum of the action $S[\dL]$, that is, that there is no deeper minimum. For positive $y$ the action $S[\dL]$ written in eq.(\ref{S1}) satisfies the lower bound:
\beq
y \geq 0 : \;\; S[\dL] \geq \frac{\sigma^2(R)}{2} \; \dL . \DL^{-1} . \dL \geq 0
\label{Smin1}
\eeq
since the density $\rho_R$ is positive. Moreover, the kernel $\DL^{-1}$ is positive definite, as shown by eq.(\ref{Dl2}). Hence the action $S[\dL]$ goes to $+\infty$ for large linear density fields $|\dL| \rightarrow +\infty$. Since we obtained only one spherical saddle-point in Sect.\ref{Spherical saddle-point} (there is only one solution $\tau(y)$ of eq.(\ref{Sy1}) for positive $y$, see also \cite{paper2}) we can conclude that this saddle-point is also the global minimum of the action $S[\dL]$ restricted to spherical states $\dL$. Then, we need to show that there is no deeper minimum realized for a non-spherical state $\dL$. Unfortunately, since we do not know the explicit form of the functional $\rho_R[\dL]$ we cannot prove this in a rigorous fashion. Hence we shall simply assume that the spherical saddle-point derived in Sect.\ref{Spherical saddle-point} is indeed the deeper minimum of the action. Note that this assumption is actually quite reasonable. Indeed, for a given ``magnitude'' of the linear density field $\dL$ we can expect spherical states to be the most efficient to build very underdense spherical regions.

The second point we need to check in order to validate the steepest-descent method is to show that the Gaussian integration we performed in eq.(\ref{psib3}) is justified. In other words, we must show that in the limit $y \rightarrow +\infty$ (which also corresponds to $\rho_R \rightarrow 0$) which we study here, the action $S[\dL]$ in the path-integral (\ref{psib2}) can be replaced by its Taylor expansion up to second-order. Unfortunately, this is a rather difficult task since we do not know explicitly the functional $\rho_R[\dL]$. Therefore, it is not easy to estimate the high-order derivatives of the action $S[\dL]$ defined in eq.(\ref{S1}). Moreover, as noticed in \cite{paper2} it is not obvious that one could use standard perturbation theory around the spherical saddle-point $\dL$ to estimate these higher order derivatives.Indeed, one can check that using the form of the $n-$order term $\delta^{(n)}(\bx)$ obtained from the expansion of the non-linear density field as a power-series over the linear density field $\dL$ leads to divergent quantities (see also \cite{paper5}). This is similar to the usual divergences one encounters in high-order terms (``loop corrections'') obtained from such expansions to compute the non-linear two-point correlation (e.g., \cite{Scoc1}). Therefore, we shall only show that the steepest-descent method is justified with respect to the integration over the one-dimensional variable $\tau$. That is, we show that it is valid for the quantity:
\beq
\psib_1(y) \equiv \int_{-\infty}^{+\infty} \frac{\d\tau}{\sqrt{2\pi}\sigma} \; e^{-S_1(\tau)/\sigma^2} ,
\label{psi11}
\eeq
with:
\beq
S_1(\tau) \equiv y \; \cGr(\tau) + \frac{\tau^2}{2} .
\label{S11}
\eeq
The ordinary integral (\ref{psi11}) is a simplified version of the path-integral (\ref{psib2}), where we replace the infinite-dimensional variable $\dL(\bx)$ by the real variable $\tau$. It also exhibits a saddle-point $\tau_c$ which is still given by eq.(\ref{Sy1}) and the value of the exponent at this point is again given by eq.(\ref{Sy1}) with $S_1(\tau_c) = S_y$. For large $y$ we can use the asymptotic behaviour (\ref{asytau1}) and the saddle-point $\tau_c$ is given by eq.(\ref{tauy1}). Then, making the change of variable $\tau=\tau_c x$ we can write eq.(\ref{psi11}) as:
\beq
\psib_1(y) \simeq \tau_c \int_0^{\infty} \frac{\d x}{\sqrt{2\pi}\sigma} \; e^{-\tau_c^2 \left[ \frac{1-n}{6} x^{-6/(1-n)} + \frac{1}{2} x^2 \right]/\sigma^2 }
\label{psi12}
\eeq
where we restricted the integration over $x>0$ since for large positive $y$ the integral (\ref{psi11}) is dominated by large positive $\tau$. Then, it is clear from eq.(\ref{psi12}) that the steepest-descent method yields exact results in the limit $y \rightarrow +\infty$, which corresponds to $\tau_c \rightarrow +\infty$. We can expect this behaviour to remain valid (for the most part) for the path-integral (\ref{psib2}). Therefore, we shall assume in the following that the steepest-descent method is indeed justified.

\subsection{Asymptotic form of the generating function}
\label{Asymptotic form of the generating function}

We can now evaluate the generating function $\psib(y)$ obtained in eq.(\ref{psib3}). This expression can also be written:
\beq
\psib(y) = \cD^{-1/2} e^{- S_y /\sigma^2(R)} \;\; \mbox{with} \;\; \cD = \Det(\DL.M_y) .
\label{psib4}
\eeq
Using eq.(\ref{Sy1}) and eq.(\ref{asytau1}) this yields the asymptotic behaviour for large $y$:
\beq
y \rightarrow +\infty : \; S_y = a \; y^{1-\om}
\label{Sy3}
\eeq
for the minimum of the action $S$, with:
\beq
a = \frac{4-n}{6} \left( \frac{6}{1-n} \right)^{(1-n)/(4-n)} \left( \frac{20}{27} \right)^{-6/(4-n)}
\label{a1}
\eeq
and:
\beq
\om = \frac{3}{4-n} .
\label{om1}
\eeq
Next, we need to evaluate the determinant $\cD$. However, since it involves the second-order derivative of the functional $\rho_R[\dL]$, see eq.(\ref{M1}), we do not know its explicit expression. Therefore, as in \cite{paper2} (App.B) we shall use the analogy with the simple integral (\ref{psi11}). The steepest-descent method yields for $\psib_1(y)$ the expression:
\beq
\psib_1(y) \simeq \frac{1}{\sqrt{1+y\cGr''(\tau)}} \; e^{-S_y/\sigma^2}
\label{psi13}
\eeq
where $\tau$ is the saddle-point given by eq.(\ref{Sy1}). Hence in the case of the path-integral (\ref{psib2}) we simply have the substitution $1+y\cGr''(\tau) \rightarrow \cD$. Thus, in a first step we may approximate the determinant $\cD$ by the simple quantity $1+y\cGr''(\tau)$. However, as noticed in \cite{paper2} this analogy does not take into account a physical process which is specific to the non-local problem of large-scale structure formation: the stronger expansion of underdense regions. Indeed, the ordinary integral (\ref{psi11}) only applies to a local physics. On the other hand, while their density contrast decreases towards $-1$ underdense regions also depart from the mean background expansion and they actually grow in comoving coordinates. In fact, the volume they occupy is larger by a factor $(R/R_L)^3=\rho_R^{-1}$ relative to the initial comoving value. This increases the pdf $\cP(\rho_R)$ as well as the average $\psib(y) = \lag e^{-y\rho_R/\sigma^2} \rag$ by the same factor. Therefore, we approximate the generating function $\psib(y)$ by:
\beq
y \rightarrow +\infty : \;\; \psib(y) \simeq \frac{1}{\rho_R(\tau)} \; \frac{1}{\sqrt{1+y\cGr''(\tau)}} \; e^{-S_y/\sigma^2(R)} .
\label{psib5}
\eeq
Here the overdensity $\rho_R$ is given by eq.(\ref{G1}). It is the overdensity associated with the saddle-point $\tau$. Note that the factor $\rho_R^{-1}$ corresponds to a non-local physics. Indeed, as the underdense bubble grows in comoving coordinates it gobbles neighbouring regions whose density becomes governed by the properties of this initially remote underdense patch.

Note that we can actually see the factor $1/\rho_R$ appear in the path-integral (\ref{psib2}) through the following discussion. When we translate the spherical saddle-point $\dL(\bx)$ obtained in Sect.\ref{Spherical saddle-point} by a vector $\br$ we do not change the term $(\dL.\DL^{-1}.\dL)$ in the action (\ref{S1}), see eq.(\ref{Dl2}), since a translation only yields a phase shift to the Fourier components $\dL(\bk)$ (this is related to the fact that the initial conditions are homogeneous). Moreover, the overdensity $\rho_R[\dL]$ is not significantly changed as long as the displaced ``bubble'' of radius $R$ shows some overlap with the spherical cell of radius $R$ centered on the origin. This holds as long as $r \la R$. Thus, this yields a set of linear states $\delta_L^{(\br)}(\bx)$ over which the action $S[\dL]$ is almost degenerate. Therefore, we must sum up the contributions of these various states (this is similar to the usual case of degenerate saddle-points). If we discretize the comoving coordinates $\bx$ by a grid of step $\Delta x$ we obtain a number of such states which scales as $(R/\Delta x)^3$ (since they are translated from the origin by a length $r$ of order of or smaller than $R$). Next, we normalize to the local physics where each region follows the mean expansion of the universe, which yields a factor $(R/R_L)^3=1/\rho_R$ (the discrete step $\Delta x$ eventually disappears as it should). Therefore, we naturally recover the prefactor $1/\rho_R$.

This mechanism also shows that we could have some additional deviations from the prefactor written in eq.(\ref{psib5}) if the dependence of the action on the magnitude of $\dL$ is not the same along all directions (i.e. there are almost flat or very sharp directions). However, this appears rather unlikely. Indeed, as shown in eq.(\ref{psi12}) the variation of the action around the saddle-point is governed at the same order (in the simple case) by the functional $\rho_R[\dL]$ and by the simple quadratic term $(\dL.\DL^{-1}.\dL)$. This last term only remains constant for translations (treated above, which yield the factor $1/\rho_R$) and for rotations. However, this last symmetry only gives a constant factor $4 \pi$ which is absorbed into the normalization of the path-integral. Therefore, the quadratic term $(\dL.\DL^{-1}.\dL)$ should prevent any new ``flat direction''. On the other hand, there might exist ``sharp'' directions if the density $\rho_R$ were strongly unstable with respect to non-spherical perturbations, for instance. However, this is not the case here since the dynamics is actually stable around the saddle-point. Indeed, it is well-known that in the similar case of the expansion of low-density universes (i.e. with $\Om < 1$) density perturbations do not grow (i.e. the linear growing mode $D_+(t)$ saturates as soon as $\Om \la 0.2$), see \cite{Peeb1}. However, we can clearly expect that an exact calculation would give a multiplicative numerical factor of order unity with respect to eq.(\ref{psib5}). In any case, note that the exponential term is exact (for $-3<n \leq -1$) since it only depends on the value of the action at the saddle-point.

\subsection{Low-density tail of the pdf $\cP(\rho_R)$}
\label{Low-density tail of the pdf}

From the generating function $\psib(y)$ obtained in eq.(\ref{psib5}) we can now derive the low-density tail of the pdf $\cP(\rho_R)$. Indeed, from the inverse Laplace transform (\ref{P1}) and the definition (\ref{psib1}) we can write:
\beq
\cP(\rho_R) = \inta \frac{\d y}{2\pi i \sigma^2(R)} \; e^{y \rho_R/\sigma^2(R)} \; \psib(y) .
\label{P2}
\eeq
Using eq.(\ref{psib5}) this yields:
\beq
\cP(\rho_R) \simeq \inta \frac{\d y}{2\pi i \sigma^2} \; \frac{1}{\cGr(\tau) \sqrt{1+y\cGr''(\tau)}} \; e^{[y \rho_R -S_y]/\sigma^2}
\label{P3}
\eeq
Then, in the limit $\rho_R \rightarrow 0$ we can evaluate this integral by an ordinary steepest-descent method which gives:
\beq
\cP(\rho_R) \simeq \frac{1}{\sqrt{2\pi}\sigma} \; \frac{1}{\rho_R \sqrt{1+y\cGr''(\tau)}} \; \frac{1}{\sqrt{-S_y''}} \; e^{-\tau^2/(2\sigma^2)} .
\label{P4}
\eeq
Here the variable $\tau$ is again given by $\cGr(\tau) = \rho_R$ while $S_y''$ is the second-derivative with respect to $y$ of the value of the action $S_y$ at the saddle-point associated with $\tau$. From eq.(\ref{Sy1}) we get:
\beq
S_y'' = \cGr'(\tau) \frac{\d\tau}{\d y} , \;\; 1+y\cGr''(\tau)= - \cGr'(\tau) \frac{\d y}{\d\tau} ,
\label{Syd1}
\eeq
so that eq.(\ref{P4}) also writes:
\beq
\cP(\rho_R) \simeq \frac{1}{\sqrt{2\pi}\sigma} \; \frac{1}{\rho_R} \; \frac{1}{|\cGr'(\tau)|} \; e^{-\tau^2/(2\sigma^2)} .
\label{P5}
\eeq
Then, using eq.(\ref{asytau1}) we obtain:
\beq
\cP(\rho_R) \simeq \frac{1}{\sqrt{2\pi}\sigma} \; \frac{1-n}{6} \; \frac{27}{20} \; \rho_R^{\frac{n-13}{6}} \; e^{-(\frac{27}{20})^2 \rho_R^{-(1-n)/3} /(2\sigma^2)} .
\label{P6}
\eeq
The asymptotic expressions (\ref{P5}) and (\ref{P6}) hold for very rare underdensities, beyond the cutoff $\tau_v$ or $\rho_v$ of the pdf $\cP(\rho_R)$ (``v'' for voids). This characteristic underdensity is set by the condition $\tau \sim \sigma(R)$, see eq.(\ref{P5}). In the quasi-linear regime we have $\tau \simeq -\dR$, see eq.(\ref{tau1}) and \cite{paper2}, which gives $\delta_v \sim -\sigma(R)$ as expected. In the non-linear regime where $\sigma(R) \gg 1$ we have $\tau_v \gg 1$ so that we can use the asymptotic behaviour (\ref{asytau1}) which yields $\rho_v \sim \sigma(R)^{-6/(1-n)}$. Therefore, the typical overdensity $\rho_v$ of voids is given by:
\beq
\sigma \ll 1 : \rho_v = 1 - \sigma(R) , \;\;\;  \sigma \gg 1 : \rho_v = \sigma(R)^{-6/(1-n)} .
\label{rhov1}
\eeq
In the non-linear regime this yields:
\beq
\sigma(R) \gg 1 : \; \rho_v = \sigma(R)^{-6/(1-n)} \propto R^{3(n+3)/(1-n)} .
\label{rhov2}
\eeq
Thus, the expressions (\ref{P5}) and (\ref{P6}) provide the asymptotic behaviour of the low-density tail of the pdf $\cP(\rho_R)$. They apply to $\rho_R \ll \rho_v$. From eq.(\ref{asytau1}) and eq.(\ref{tauy1}) we see that the density $\rho_v$ corresponds to the value $y_v$ of the Laplace variable $y$, with:
\beq
\sigma \gg 1 : y_v = \rho_v^{-(4-n)/3} = \sigma^{(8-2n)/(1-n)} .
\label{yv1}
\eeq
In particular, the asymptotic form (\ref{psib5}) of the generating function $\psib(y)$ actually applies to $y \gg y_v$.

Here, we must stress that the results (\ref{P5}) and (\ref{P6}) have been directly derived from the equations of motion. For $n \leq -1$ the exponential in eq.(\ref{P6}) is exact but the prefactor is only approximate because of the approximation (\ref{psib5}) we used for the determinant $\cD$. For $n >-1$ the normalization factor in the exponential is also approximate because we did not use the exact saddle-point, as discussed in Sect.\ref{Spherical saddle-point}. However, the characteristic exponent $\om$ in eq.(\ref{Sy3}) and the power $\rho_R^{-(1-n)/3}$ within the exponential cutoff in eq.(\ref{P6}) should still be exact. Besides, we recall that eq.(\ref{P6}) holds for arbitrary values of the variance $\sigma$, provided $\rho_R \ll \rho_v$. Indeed, we only used the limit of very underdense regions (rare voids) and the actual value of $\sigma$ was irrelevant in the calculation. In fact, the normalization of the power-spectrum does not even appear in the characteristic action $S[\dL]$ defined in eq.(\ref{S1}) which describes the physics of our system ! Thus, our result (\ref{P6}) is valid from the quasi-linear regime up to the highly non-linear regime. The influence of $\sigma$ only shows up in the constraint $\rho_R \ll \rho_v$, as shown by eq.(\ref{rhov1}). Indeed, it is clear that as time goes on and gravitational clustering builds up the characteristic overdensity $\rho_R$ of typical voids evolves. That is, the cutoff at low densities of the pdf $\cP(\rho_R)$ is steadily pushed towards lower densities as gravitational clustering proceeds.

The quasi-linear regime was already studied on its own in \cite{paper2}. In that previous work, we investigated the limit $\sigma \rightarrow 0$ which gave the pdf $\cP(\rho_R)$ (or $\cP(\dR)$) for all density contrasts $\dR$ in the quasi-linear regime. Indeed, in this limit any finite density contrast becomes a rare event as soon as $\sigma \ll |\dR|$. Since the calculation involved the same spherical saddle-point (\ref{col2}) described in Sect.\ref{Spherical saddle-point} the expression (\ref{P6}) is consistent with the results obtained in that previous paper. Of course, the main interest of eq.(\ref{P6}) is that it also applies to the non-linear regime $\sigma > 1$. Indeed, this regime is more difficult to handle and very few rigorous results were known so far. Moreover, in order to build significant underdensities one must be at least in the mildly non-linear regime $\sigma \ga 1$.

\subsection{Comparison with previous works}
\label{Comparison with previous works}

\subsubsection{Perturbative methods}
\label{Perturbative methods}

We can note that the approach described in the previous sections bears some similarity with a perturbative study developed in \cite{Ber1}. This author investigated the mean non-linear dynamics in the limit of rare events (i.e. large density fluctuations) through a perturbative method which assumes that the non-linear density field can be written as a perturbative expansion over the linear density field. Note that this problem is not obvious a priori. Indeed, although the mean density profile, with the constraint that the average linear density contrast within the radius $R_L$ is some fixed value $\delta_{L,R_L}$, obeys eq.(\ref{col2}) (i.e. exactly the profile of our saddle-point), it does not follow that the mean non-linear profile should be given by the non-linear evolution of the mean linear profile (i.e. these two operations may not commute). However, \cite{Ber1} noticed from a perturbative treatment that this is actually the case as one eventually recovers the usual spherical dynamics. This agrees with our results (\ref{P5}) and (\ref{P6}). In fact, our approach provides a simple explanation for this feature. This peculiar spherical solution of the dynamics is actually a saddle-point of the action $S[\dL]$ so that it governs the tails of the pdf $\cP(\rho_R)$. Note also that our method is much more intuitive and simpler as it clearly reveals the underlying physics.

Next, we stress that our method is non-perturbative and it yields exact results (as long as one can identify the exact minimum of the action). By contrast, perturbative methods are based on the expansion of the non-linear density field over the growing linear density field, which is then plugged into the hydrodynamical approximation of the equations of motion. Therefore, such approaches do not provide complete proofs since the perturbative expansions should diverge. Moreover, they cannot go beyond shell-crossing. Thus, \cite{Ber1} noticed that for power-spectra with $n\geq -1$ the first correction (i.e. subleading term) obtained by the perturbative approach diverges. This implies that the perturbative method fails for $n\geq -1$. This feature can easily be understood from the discussion given in Sect.\ref{Spherical saddle-point}. Indeed, we noticed that for $n>-1$ the spherical saddle-point of the action experiences some shell-crossing for $R_L' \sim R_L$. Therefore, it cannot be obtained by perturbative means and all perturbative methods must diverge. Nevertheless, this does not invalidate our steepest-descent method which is essentially non-perturbative. It merely means that it is more difficult to obtain an analytical expression for the exact minimum of the action $S[\dL]$. Besides, as described in Sect.\ref{Spherical saddle-point} our approach provides a very convenient way to obtain approximate results in such cases. We simply need to use an approximation for the spherical saddle-point. As argued in Sect.\ref{Spherical saddle-point} we can expect this procedure to provide very good results for the case of rare underdensities since even for $n>-1$ shell-crossing only involves a limited range of radii along the density profile of the saddle-point. In particular, we expect that all characteristic exponents (e.g. the power $\rho_R^{-(1-n)/3}$ in the exponential term in eq.(\ref{P6})) should remain exact.

Note that this feature for $-1<n<1$ clearly shows once more that perturbative results should be viewed with caution. Indeed, at leading-order the perturbative approach yields the usual spherical dynamics, as in eq.(\ref{P6}). However, as we explained in Sect.\ref{Spherical saddle-point} this does not correspond to the exact spherical saddle-point which means that the leading-order behaviour of the pdf $\cP(\rho_R)$, or of the generating function $\psib(y)$, is not given by this simple expression. Indeed, we can expect the actual minimum of the action $S[\dL]$ to be slightly different from the value computed from eq.(\ref{col1}), which translates into a slightly different value for the numerical factor $a$ in eq.(\ref{Sy3}). Therefore, the divergence of the subleading terms actually leads to a correction to the finite leading term derived from the perturbative method. Then, we note that for hierarchical scenarios all perturbative series actually diverge because on small scales the density field is always a non-perturbative quantity (see \cite{paper1} and \cite{paper5}). Moreover, it is well-known that in standard perturbative expansions one actually encounters divergent quantities beyond some finite order (``loop corrections'', e.g. \cite{Scoc1}). As a consequence, there can be no guarantee that perturbative results (even though finite and restricted to leading-order terms) make sense. The only way to obtain firm results is to use non-perturbative methods which can overcome (at least in principle) these problems. The goal of our previous work (\cite{paper2}) and of the present article is precisely to develop such tools.

Finally, since we shall investigate the high-density tail of the pdf in Sect.\ref{Rare overdensities} we note here that perturbative methods as in \cite{Ber1} are restricted to the early non-linear stages of the dynamics before shell-crossing occurs. By contrast, the high overdensities we shall study correspond to non-linear objects which have already virialized and where shell-crossing plays a key role.

\subsubsection{Spherical model}
\label{Spherical model}

As in \cite{paper2}, we note that the expression (\ref{P5}) can actually be recovered from a very simple spherical model, detailed for instance in \cite{Val2}. This phenomenological model rests on the approximation:
\beq
\int_{\dR}^{\infty} \d\delta \; (1+\delta) \cP(\delta) \simeq \int_{\delta_{L,R_L}}^{\infty} \d\dL \; \cP_L(\dL) .
\label{spher1}
\eeq
Here $\cP_L(\dL)$ is the linear pdf of the linear density contrast $\dL$ within a spherical cell. This relation merely states that the fraction of matter contained within spherical cells of radius $R$ and non-linear density contrast larger than $\dR$ is approximately equal to the fraction of matter which is enclosed within spherical cells of Lagrangian radius $R_L$ and linear density contrast greater than $\delta_{L,R_L}$. Here $R_L$ and $\delta_{L,R_L}$ are related to $R$ and $\dR$ by the usual spherical dynamics, as in eq.(\ref{F1}). Note that this is very close to the Press-Schechter prescription without the factor 2, see \cite{PS}. Next, we note that the linear pdf $\cP_L(\dL)$ at scale $R$ exhibits a simple scaling over the variable $\nu$ through:
\beq
\cP_L(\dL) \; \d\dL = \cP_L^{(\nu)}(\nu) \; \d\nu \;\;\; \mbox{with} \;\;\; \nu = \frac{\dL}{\sigma(R)}
\label{PLnu1}
\eeq
and:
\beq
\cP_L^{(\nu)}(\nu) \equiv \frac{1}{\sqrt{2\pi}} \; e^{-\nu^2/2} .
\label{Pnu1}
\eeq
Substituting eq.(\ref{PLnu1}) into eq.(\ref{spher1}) and differentiating with respect to $\dR$ we obtain:
\beq
\cP_s(\dR) = \frac{1}{1+\dR} \; \frac{1}{\sqrt{2\pi}} \; \frac{\d\nu}{\d\dR} \; e^{-\nu^2/2}
\label{Psnu1}
\eeq
with:
\beq
\nu = \frac{\delta_{L,R_L}}{\sigma(R_L)} = - \frac{\tau}{\sigma(R)} .
\label{nutau}
\eeq
Here the subscript ``s'' refers to the ``spherical'' model. In eq.(\ref{nutau}) we introduced the variable $\tau$ defined in eq.(\ref{tau1}). Moreover, using the function $\cGr(\tau)$ introduced in eq.(\ref{G1}) we have:
\beq
\frac{\d\nu}{\d\dR} = - \frac{1}{\sigma(R)} \; \frac{\d\tau}{\d\dR} = \frac{1}{\sigma(R)} \; \frac{1}{|\cGr'(\tau)|} .
\label{dnuddR}
\eeq
Then, substituting eq.(\ref{dnuddR}) into eq.(\ref{Psnu1}) we recover the expression (\ref{P5}). Thus, for low densities $\rho_R \ll \rho_v$ the spherical dynamics correctly describes the leading order behaviour of the pdf $\cP(\rho_R)$. This is not surprising: it is simply due to the fact that the saddle-point of the action $S[\dL]$ obtained in Sect.\ref{Spherical saddle-point} is spherically symmetric. This holds because the initial conditions are homogeneous and isotropic and we study the statistics of the mean overdensity $\rho_R$ over a spherical cell of radius $R$, centered on the origin (for instance), which preserves the spherical symmetry of the problem. Note however that the determinant $\cD$ which appears in the prefactor of the generating function $\psib(y)$ in eq.(\ref{psib4}) takes into account the deviations of the initial conditions from spherical symmetry (at leading order in the limit $\rho_R \rightarrow 0$). As discussed in Sect.\ref{Asymptotic form of the generating function}, we can expect the exact Gaussian integration over the non-spherical density fluctuations $\dL$ around the saddle-point to give a multiplicative numerical factor of order unity with respect to eq.(\ref{P5}). Moreover, as shown in Sect.\ref{Spherical saddle-point} for $-1<n<1$ the spherical model only yields an approximation for the exponential cutoff of the low-density tail since the saddle-point given by eq.(\ref{col1}) is no longer exact.

As already noticed in \cite{Val2}, the overdensity $\rho_v$ obtained in eq.(\ref{rhov2}) corresponds to density fluctuations which would occupy all the volume of the universe at the time of interest. This is the typical density of voids which fill almost all the volume of the universe in the non-linear regime. It is clear that for higher densities the deviations from the spherical dynamics and the effects of shell-crossing play a key-role and they must be taken into account.

\subsubsection{Non-linear scaling model}
\label{Non-linear scaling model}

Here, we point out that our results (\ref{Sy3}) and (\ref{P6}) are reminiscent of the prediction of a non-linear hierarchical ansatz investigated in \cite{Bal1}. This model is based on the assumption that in the highly non-linear regime ($\sigma \gg 1$) the non-linear many-body correlation functions $\xi_q(\bx_1,..,\bx_q;a)$ obey the scaling law:
\beq
\xi_q(\lambda \bx_1,..,\lambda \bx_q;a) = a^{3(q-1)} \; \lambda^{-\gam(q-1)} \; \hat{\xi}_q(\bx_1,..,\bx_q)
\label{scal1}
\eeq
for arbitrary $\lambda >0$ and any time. Here $a(t)$ is the scale-factor while $\gam$ is the slope of the non-linear two-point correlation function $\xi$. This scaling law can be derived from the stable-clustering assumption (\cite{Peeb1}). In this case, for a power-law power-spectrum $P(k) \propto k^n$ we have:
\beq
\gam = \frac{3(3+n)}{5+n} .
\label{gam1}
\eeq
Then, it is convenient to introduce the quantities:
\beq
S_q \equiv \frac{\xib_q}{\xib^{\;q-1}} \;\; \mbox{with} \;\; \xib_q(R) \equiv \int_V \frac{\d \bx_1 .. \d \bx_q}{V^q} \; \xi_q(\bx_1,..,\bx_q) 
\label{Sq1}
\eeq
where we note $\xib=\xib_2$. Thus, the parameters $S_q$ yield the cumulants $\lag \rho_R^q \rag_c \equiv \xib_q$ (with $\xib_1 \equiv 1$). Besides, the scaling laws (\ref{scal1}) imply that the coefficients $S_q$ do not depend on scale nor on time. Next, the Laplace transform $\psi(y)$ of the pdf defined in eq.(\ref{psi1}) is related to these cumulants by the standard property (see any textbook on probability theory):
\beq
\ln[\psi(y)] = \sum_{q=1}^{\infty} \frac{(-1)^q}{q!} \; \lag \rho_R^q \rag_c \; y^q .
\label{cum1}
\eeq
Using eq.(\ref{Sq1}) this yields:
\beq
\psi(y) = \psit(y \xib) \;\; \mbox{with} \;\; \psit(y) \equiv e^{-\phit(y)/\xib}
\label{psit1}
\eeq
and:
\beq
\phit(y) \equiv \sum_{q=1}^{\infty} \frac{(-1)^{q-1}}{q!} \; S_q \; y^q .
\label{phit1}
\eeq
Note that within this model the generating function $\phit(y)$ is scale and time independent. Besides, it provides the pdf $\Pt(\rho_R)$ through eq.(\ref{P1}) which reads:
\beq
\Pt(\rho_R) = \inta \frac{\d y}{2\pi i \xib} \; e^{[y \rho_R - \phit(y)]/\xib}
\label{Pt1}
\eeq
where the tilde ``$\sim$'' refers to the hierarchical ansatz. As argued in \cite{Bal1}, it is natural to expect a power-law asymptotic behaviour at large $y$:
\beq
y \rightarrow +\infty : \; \phit(y) \simeq \at \; y^{1-\omt} \;\; \mbox{with} \;\; \at>0 , \; 0 \leq \omt \leq 1 .
\label{phit2}
\eeq
Note that this is similar to our results (\ref{psib4}) and (\ref{Sy3}) for the generating function $\psib(y)$. Apart for the factor $\cD$ in eq.(\ref{psib4}), the only difference is that within this hierarchical ansatz the linear variance $\sigma^2$ must be replaced by its non-linear counterpart $\xib$. From eq.(\ref{Pt1}) and eq.(\ref{phit2}) one obtains for small overdensities $\rho_R \ll \xib$ the behaviour (\cite{Bal1}):
\beq
\rho_R \ll \xib : \;\; \Pt(\rho_R) = \at^{-1/(1-\omt)} \; \xib^{\;\omt/(1-\omt)} \; g_{\omt}(z)
\label{Pt2}
\eeq
where we introduced the variable $z$ defined by:
\beq
z \equiv \at^{-1/(1-\omt)} \; \xib^{\;\omt/(1-\omt)} \; \rho_R
\label{z1}
\eeq
and the function $g_{\omt}(z)$ given by:
\beq
g_{\omt}(z) \equiv \inta \frac{\d t}{2\pi i} \; e^{z t - t^{1-\omt}} .
\label{gz1}
\eeq
The function $g_{\omt}(z)$ exhibits a sharp cutoff for small $z$ which can be computed by an ordinary steepest-descent method. This eventually yields (\cite{Bal1}):
\beqa
z \ll 1 : \Pt(\rho_R) & = & \at^{-1/(1-\omt)} \; \xib^{\;\omt/(1-\omt)} \; \sqrt{\frac{(1-\omt)^{1/\omt}}{2\pi \omt}} \nonumber \\ & & \times \; z^{-(1+\omt)/(2\omt)} \; e^{-\omt [z/(1-\omt)]^{-(1-\omt)/\omt}} .
\label{Pt3}
\eeqa
At first sight, this cutoff looks similar to eq.(\ref{P6}). First, we note that we can make the typical void density $\rhot_v$ implied by eq.(\ref{Pt2}) to coincide with our result (\ref{rhov2}). Indeed, the value of the parameters $\at$ and $\omt$ is not predicted by the hierarchical ansatz (\ref{scal1}). The low-density cutoff of the pdf $\Pt(\rho_R)$ is given by $z \sim 1$, which yields for the typical overdensity $\rhot_v$ of voids:
\beq
\sigma(R) \gg 1 : \; \rhot_v = \xib^{\;-\omt/(1-\omt)} \propto R^{\gam \omt/(1-\omt)} .
\label{rhotv1}
\eeq
Using eq.(\ref{gam1}), the comparison with our result (\ref{rhov2}) for $\rho_v$ gives:
\beq
\omt = \frac{5+n}{6} .
\label{omt1}
\eeq
Here, it is interesting to note that the value (\ref{omt1}) for $\omt$ was already obtained in \cite{Val1} from the stable-clustering ansatz coupled with the spherical collapse model. Let us recall briefly the main properties of this model. As in Sect.\ref{Spherical model}, it is based on the approximation (\ref{spher1}) which is used to relate the non-linear density field $\delta(\bx)$ to its linear counterpart $\dL(\bx)$. Then, in \cite{Val1} we used eq.(\ref{spher1}) to ``derive'' the pdf $\cP(\rho_R)$ in the non-linear regime. In particular, we considered collapsed objects which have already virialized. Then, within the framework of the stable-clustering ansatz we have $\rho_R \propto \rhob^{-1} \propto a^3$ while the linear density contrast obeys $\dL(M) \propto a$ (in a critical density universe) which yields $\rho_R \propto \dL^3$ and $\cF(\dL) \propto \dL^3$. Next, substituting this behaviour into eq.(\ref{spher1}) we obtain:
\beq
\rhot_v \ll \rho_R \ll \xib : \;\; \Pt(\rho_R) \propto \frac{1}{\xib^{\;2}} \; \left(\frac{\rho_R}{\xib}\right)^{\omt-2} .
\label{Pt4}
\eeq
where the exponent $\omt$ is given by eq.(\ref{omt1}) and the low-density cutoff is given by eq.(\ref{rhotv1}). Note that the power-law behaviour (\ref{Pt4}) agrees with the scaling (\ref{Pt2}) over the range $\rhot_v \ll \rho_R \ll \xib$ where both formulae overlap. Indeed, for large values of $z$ the function $g_{\omt}(z)$ defined in eq.(\ref{gz1}) obeys the asymptotic behaviour (\cite{Bal1}):
\beq
z \gg 1 : \;\; g_{\omt}(z) \simeq \frac{1-\omt}{\Gamma(\omt)} \; z^{\omt-2} .
\label{gz2}
\eeq
This is obtained by expanding the term $-t^{1-\omt}$ in the exponent in eq.(\ref{gz1}) since for $z \gg 1$ only small values of $t$ contribute to the integral. Here, we must point out that this line of reasoning involves virialized objects which satisfy $\rho_R \gg \rhot_v$, whence the lower bound for the range of validity of eq.(\ref{Pt4}). In particular, at this stage stable-clustering does not imply the scaling (\ref{Pt2}) for $\rho_R \ll \rhot_v$. In order to obtain eq.(\ref{Pt2}) down to $\rho_R \rightarrow 0$ one needs to assume that the power-law behaviour (\ref{phit2}) extends up to $y \rightarrow +\infty$ at all times and scales (the function $\phit(y)$ is scale and time-independent in this framework), as assumed in \cite{Bal1}. However, this is clearly inconsistent with our rigorous results (\ref{P5}) and (\ref{P6}). Indeed, using eq.(\ref{omt1}) we obtain $(1-\omt)/\omt=(1-n)/(5+n)$. This implies from eq.(\ref{Pt3}) that the pdf $\Pt(\rho_R)$ exhibits a low-density tail of the form $\Pt(\rho_R) \sim e^{-(\rho_R/\rhot_v)^{-(1-n)/(5+n)}}$ which disagrees with eq.(\ref{P6}). Thus, our study explicitly shows that the non-linear hierarchical ansatz (\ref{scal1}) does not describe the pdf $\cP(\rho_R)$ for rare underdensities $\rho_R \ll \rho_v$. In other words, the very low density tail of the pdf cannot be derived from the stable-clustering ansatz. 

Actually, this is not surprising in view of the physics which lies behind the derivation performed in Sect.\ref{Spherical saddle-point} and Sect.\ref{Generating function}. Indeed, we have shown that the low density tail of the pdf (i.e. $\rho_R \ll \rho_v$) is governed by the dynamics of spherical very rare ``low-density bubbles'' which are still expanding. In particular, in this asymptotic regime the expansion of the outer shells is almost ``free'', that is their physical radius grows as $R \propto t$ which implies $\rho_R \propto t^{-1} \propto \dL^{-3/2}$ (if $\Om=1$) in agreement with eq.(\ref{asymp1}). Indeed, the gravitational pull from the inner regions becomes negligible. Note also that virialization processes do not show up at all. Therefore, we could have expected the stable-clustering ansatz (\ref{scal1}) to be irrelevant to the behaviour of the pdf $\cP(\rho_R)$ in this very low density regime. Nevertheless, the stable-clustering ansatz can be made to recover the correct void density $\rho_v$ in eq.(\ref{rhotv1}) because within this framework these underdense regions have just stopped their expansion and they ``virialize'' at the time of interest. Hence the stable-clustering assumption does not have any influence on their properties yet.

Note however that this result does not imply that the scaling laws (\ref{scal1}) are wrong. Indeed, it is clear that the many body correlation functions $\xi_q$ involved in eq.(\ref{scal1}) are dominated by the high-density regions $\rho_R \ga \xib$, which may be governed by virialization processes. Our result (\ref{P6}) merely means that the low-density tail of the pdf (i.e. $\rho_R \ll \rho_v$) depends on the detailed behaviour of the many body correlation functions $\xi_q$ which is not fully captured by the lowest-order asymptotic behaviour (\ref{scal1}). Indeed, the discrepancy between eq.(\ref{P6}) and eq.(\ref{Pt2}) means that the power-law behaviour (\ref{phit2}) only applies up to $y \la \yt_v$ at most, with:
\beq
\yt_v \equiv \rho_v^{-1/\omt}
\label{ytv1}
\eeq
as can be obtained from eq.(\ref{Pt1}). In the highly non-linear regime we have $\rho_v \rightarrow 0$ hence we get $\yt_v \rightarrow \infty$. Therefore, it is clear that the behaviour of $\phit(y)$ for $y > \yt_v$ cannot be described by the asymptotic scaling laws (\ref{scal1}), even if the latter are valid, since in the highly non-linear limit $\sigma \rightarrow +\infty$, which defines the limiting generating function $\phit(y)$ written in eq.(\ref{phit1}), this regime disappears as $\yt_v$ is repelled to $+\infty$.

\subsubsection{Numerical simulations}
\label{Numerical simulations}

Finally, our result (\ref{P6}) should be compared with numerical simulations. However, this is rather difficult since the steepest-descent method only applies to the far low-density tail of the pdf: $\rho_R \ll \rho_v$. In practice, in numerical simulations one mainly observes a sharp cutoff below some characteristic overdensity $\rho_R$ and it is not easy to measure with a good accuracy the shape of the pdf beyond this cutoff. In fact, because of discrete effects (i.e. the limited number of particles) one does not really probe the low-density tail described by eq.(\ref{P6}) in current numerical simulations. Indeed, let us note $\cP(N)$ the probability to have $N$ particles within a spherical cell of radius $R$ in a given simulation. If we assume (as is usually done) that discretization only adds a Poisson noise the pdf $\cP(N)$ is obtained from its continuous counterpart $\cP(\rho_R)$ by:
\beq
\cP(N) = \int_0^{\infty} \d\rho_R \; \cP(\rho_R) \; \frac{(\rho_R \Nb)^N}{N !} \; e^{-\rho_R \Nb} .
\label{PN1}
\eeq
The kernel in the integrand in eq.(\ref{PN1}) is simply a Poisson law and we note $\Nb$ the mean number of points within a cell of radius $R$. Next, we note $\cP_0\equiv \cP(N=0)$ the probability to find an empty cell in the simulation. Then, from eq.(\ref{PN1}) and the inverse Laplace transform (\ref{P2}) a straightforward integration over $\rho_R$ and next over $y$ yields:
\beq
\cP_0 = \psib( \Nb \sigma^2) .
\label{P01}
\eeq
Therefore, we see that $\cP_0$ probes the generating function $\psib(y)$ at $y_0=\Nb \sigma^2$. Moreover, it is clear that the numerical simulation cannot probe the continuous generating function $\psib(y)$ to larger $y$, that is the pdf $\cP(\rho_R)$ to smaller densities. Thus, the simulation only tests the low-density tail (\ref{P6}) if this value $y_0$ is much larger than the value $y_v$ obtained in eq.(\ref{yv1}) which marks the onset of the regime associated with these very underdense regions, in the non-linear regime $\sigma \gg 1$. Using eq.(\ref{yv1}) this constraint reads:
\beq
\sigma \gg 1 : \;\;  y_0 \gg y_v \;\;\; \mbox{if} \;\;\; \Nb \gg \sigma^{6/(1-n)} .
\label{y0yv1}
\eeq
To check whether this condition is realized in practice we can look at the typical numbers reached in current simulations. For instance, the numerical simulations used in \cite{Lac1} contain $128^3\simeq 2 \times 10^6$ particles within a box of size $256$ Mpc (the length scale is actually arbitrary for a scale-free power-spectrum). At the end of the simulation (when the largest scales approach the non-linear regime) the value $\sigma = 10$ (not to take a too large number for the r.h.s. in eq.(\ref{y0yv1})) has only been probed in a reliable way by the scales $R \la 4$ Mpc (the scale $8$ Mpc may have reached $\sigma=10$ but it is not accurate because of finite size effects). The mean number of particles within a cell of radius $4$ Mpc is $\Nb_4\simeq 8.4$ while for $n=-2$ (the most favourable case) we have $10^{6/(1-n)}=100$. Therefore, the constraint (\ref{y0yv1}) is very far from being satisfied. Thus, we can conclude that the low-density cutoff seen in current numerical simulations is governed by discrete effects and it does not probe the actual low-density tail (\ref{P6}) of the continuous pdf $\cP(\rho_R)$. This requires a much larger number of particles.

Finally, we note that numerical simulations have been used to check the scaling model (\ref{scal1}), see for instance \cite{Col1}. To do so, one uses the fact that within this framework the probability $\cP_0$ to find an empty cell can be written (e.g., \cite{Bal1}):
\beq
\Pt_0 = e^{-\phit(\Nb \;\xib)/\xib} .
\label{Pvoid1}
\eeq
This relation can be obtained in a manner similar to the derivation of eq.(\ref{P01}). This allows one to derive the exponent $\omt$ defined in eq.(\ref{phit2}). As explained above, our results do not confirm nor invalidate these works since the regime probed in these simulations is not covered by our present study: it corresponds to ``high densities'' $\rho_R \gg \rho_v$ above the range of validity of eq.(\ref{P6}).

\section{Rare overdensities}
\label{Rare overdensities}

Finally, we investigate in this section the high-density tail of the pdf $\cP(\rho_R)$. This problem is more difficult than the study of rare voids since shell-crossing now plays a key role. In particular, we do not know the explicit analytic expression of the functional $\rho_R[\dL]$, even when we restrict ourselves to spherically symmetric states. Therefore, we did not manage to derive the asymptotic behaviour of $\cP(\rho_R)$ for large $\rho_R$ in a fully rigorous manner. Nevertheless, we shall discuss the expected properties of this high-density tail in the spirit of the steepest-descent method used in Sect.\ref{Rare underdensities}. Here, by rare overdensities we mean massive halos which have already collapsed, that is we consider highly non-linear objects where the local dynamical time (over the radius $R$) is smaller than the age of the universe. Thus, for approximately spherical halos the particles have already undergone several oscillations through the cluster.

\subsection{Spherical collapse}
\label{Spherical collapse}

Since we look for a spherical saddle-point of the action $S[\dL]$ written in eq.(\ref{S1}) we first recall in this section the non-linear dynamics of spherical states. More precisely, we consider linear density profiles of the form:
\beq
\delta_{L,R} \propto M^{-\epsilon} \propto R^{-3\epsilon} .
\label{spherL1}
\eeq
This problem was studied in \cite{Fil1} and we briefly recall below their main results. The advantage of such power-law linear density profiles is that the evolution is self-similar. Indeed, there is no characteristic length scale or time in this problem so that the dynamics is self-similar. Thus, the system at a later time is equivalent to the same system seen at a smaller scale: a rescaling in time can be absorbed by rescaling the length scales. This allows one to explicitly solve the dynamics. For instance, the case $\epsilon=1$ was studied in \cite{Bert1} from a Lagrangian point of view, where one follows the trajectory of individual particles (or spherical shells). This trajectory $r(r_i,t)$ depends on time $t$ and on the initial radius of the particle at some fixed initial time $t_i$ (or equivalently on the mass located within this spherical shell in the linear regime). Then, the self-similarity of the dynamics implies that all particles follow the same trajectory rescaled in proper units (e.g., the radius and the time of the first turn-around). Thus, the system is fully described by a function of only one variable, e.g. the time-dependence of the radius of a particle with an arbitrary initial radius, since the trajectories of all other particles can be obtained from this one by a simple time and length rescaling. Then, this function is seen to obey an ordinary integro-differential equation which yields all properties of the system. 

The behaviour of the mass shells can be described as follows. After an initial stage of expansion (when $\delta_{L,R_L} \la 1$) the particle turns around at a radius $r_{\rm ta}$ at time $t_{\rm ta}$. Next, the particle oscillates through the center of symmetry of the system. As time goes on the mass which has already turned-around increases so that the particle is buried ever more deeply within the collapsed halo. Besides, the particles enclosed within the central regions arise from a greater number of mass shells. Therefore, the density within these inner regions grows while the amplitude of the radial oscillations of a given particle declines. However, two behaviours can occur, depending on the initial slope $\epsilon$. For sharp linear density profiles with $\epsilon > 2/3$ the amplitude of the particle oscillations asymptotically stabilizes to a finite radius (which scales as its first turn-around radius) so that the actual density profile $\rho(r,t)$ becomes time-independent (in physical coordinates $\br$). This is due to the fact that the contribution of outer mass shells to the mass enclosed within a fixed physical radius $R$ is negligible. Indeed, although an increasing number of shells ``visit'' this central region as they pass through the origin the time they spend within $r<R$ becomes steadily smaller (they spend most of their time at radii $r$ of order of their large turn-around radius). Thus, as shown in \cite{Fil1} the asymptotic non-linear density profile is:
\beq
\epsilon > \frac{2}{3} : \;\; \rho(r,t) \propto r^{- 9\epsilon/(1+3\epsilon)}
\label{rhor1}
\eeq
which does not depend on time. On the other hand, for shallower linear density profiles with $\epsilon < 2/3$ the influence of the outer mass shells is no longer negligible. Then, the amplitude of the particle oscillations no longer stabilizes: it slowly decreases as the mass which has already collapsed grows. This gives rise to a density profile which undergoes an adiabatic increase:
\beq
\epsilon < \frac{2}{3} : \;\; \rho(r,t) \propto t^{(4-6\epsilon)/(9\epsilon)} \; r^{-2} .
\label{rhor2}
\eeq
Note that the radial slope no longer varies with $\epsilon$ but the time-dependence explicitly depends on $\epsilon$.

\subsection{Candidate for a spherical saddle-point}
\label{Spherical overdense saddle-point}

Now, we can apply the results recalled in the previous section to the formation of rare massive halos. Since our system is spherically symmetric we can look for a spherical saddle-point of the action $S[\dL]$. Indeed, note that a spherical state $\delta_{L,c}(\bx)$ which is a minimum of the action restricted to spherically symmetric linear density fields is automatically a saddle-point with respect to transverse directions. Indeed, assuming the functional $\rho_R[\dL]$ can be expanded as a Taylor series around this spherical point $\delta_{L,c}$ the variation of $\rho_R$ at linear order over a small perturbation $\dL=\delta_{L,c} + \Delta \dL$ is given by:
\beq
\Delta \rho_R = \int \d\bx \; \left. \frac{\delta(\dR)}{\delta(\delta_L(\bx))}\right|_{\delta_{L,c}} \; \Delta \dL(\bx) .
\label{sad1}
\eeq
Because of spherical symmetry, the first-order derivative $\delta(\dR)/\delta(\delta_L(\bx))$ at the point $\delta_{L,c}$ only depends on $|\bx|$. Therefore, the integration over angles in eq.(\ref{sad1}) vanishes for any deviation of the form $\chi(x) Y_l^m(\theta,\phi)$ with $l\neq 0$. In a similar fashion, the linear deviation $(\delta_{L,c} . \DL^{-1} . \Delta \dL)$ which arises from the second term in the action (\ref{S1}) is also zero. Therefore, the action $S[\dL]$ only shows a quadratic variation over $\Delta \dL$ for non-spherical perturbations, which implies that $\delta_{L,c}$ is also a saddle-point with respect to these non-spherical directions.

Thus, we only need consider spherical linear density fields (assuming there are no deeper non-spherical minima). However, even the restriction of the functional $\rho_R[\dL]$ to such spherical states $\dL$ is unknown since eq.(\ref{F1}) breaks down because of shell-crossing. Nevertheless, in some cases we may still approximate the spherical functional $\rho_R[\delta_{L,R''}]$ (here $R''$ is a dummy variable) by eq.(\ref{F1}). More precisely, this should provide a good approximation as long as the slope of the density profile at large radii $R_L'>R_L$ is sufficiently large, that is $\delta_{L,R_L'} \propto R_L'^{-\alpha}$ with $\alpha>2$, see eq.(\ref{spherL1}) and eq.(\ref{rhor1}). Here, the scale $R_L$ is the Lagrangian scale (i.e. mass scale) of the particles with a turn-around radius equal to $R$. Indeed, for such a steep profile we know that the mass within the radius $R$ is (up to a factor of order unity) the mass which was enclosed within this shell in the linear regime, as we recalled in Sect.\ref{Spherical collapse}. This justifies the use of eq.(\ref{F1}). As we recalled in Sect.\ref{Spherical saddle-point} this yields a spherical saddle-point which is flat within $R_L$ and which exhibits a power-law decline at large scales of the form (\ref{profil4}), see \cite{paper2} for a detailed derivation. Then, the comparison with the constraint in eq.(\ref{rhor1}) implies $n>-1$.

Thus, we see that for $n<-1$ the linear density profile is too shallow to stabilize the turn-around radius of the inner mass shells. Then, the mass enclosed within any radius in the halo is governed by the outer shells. The non-linear density profile adjusts to $\rho(r) \propto r^{-2}$ so that the mean density within the radius $R$ only depends on the scale $R_L$ which has just turned around through $\rho_R = \rho_c (R/R_L)^{-2}$, where $\rho_c$ is a simple normalization factor of order unity. In order to derive the spherical saddle-point in this case we can proceed as follows. Since in this regime the functional $\rho_R[\dL]$ only depends on the scale $R_L$ where the mean density contrast reaches the value $\delta_c$ of order unity (i.e. the largest non-linear scale) we first minimize the action $S[\dL]$ at fixed $R_L$ and finally we minimize over $R_L$. Thus, we must minimize $S[\dL]$ with the constraint:
\beq
\int_{V_L} \frac{\d\bx}{V_L} \; \dL(\bx) = \delta_c .
\label{cons1}
\eeq
As usual, this is done through the introduction of a Lagrange multiplier $\lambda$. Therefore, we need to minimize the action $S_{\lambda}$ given by:
\beqa
S_{\lambda}[\dL,R_L] & = & y \; \rho_R(R_L) + \frac{\sigma^2}{2} \; \dL . \DL^{-1} . \dL \nonumber \\ & & + \lambda \left( \int \d\bx \; \dL(\bx) \frac{\theta(x<R_L)}{V_L} - \delta_c  \right) 
\label{Slamb1}
\eeqa
where $\theta(x<R_L)$ is a top-hat with obvious notations. Minimizing $S_{\lambda}$ with respect to $\dL(\bx)$ yields:
\beq
\dL(\bx) = - \frac{\lambda}{\sigma^2} \int_{V_L} \frac{\d\bx'}{V_L} \; \DL(\bx,\bx') .
\label{Slamb2}
\eeq
Note that the linear density profile is exactly of the form (\ref{col2}). This is not surprising since in the regime relevant to Sect.\ref{Spherical saddle-point} the density $\rho_R$ does not depend either on the details of the density profile but only on the variables $R_L$ and $\delta_{L,R_L}$. Hence the spherical saddle-point we obtain for $n<-1$ is flat within the scale $R_L$ and it decreases at larger scales where it is in the linear regime.

Thus, we see that these considerations suggest two very different behaviours. For steep power-spectra with $n>-1$ the high-density tail of the pdf $\cP(\dR)$ at scale $R$ would be related to large density fluctuations over the Lagrangian scale $R_L$ associated with the Eulerian scale $R$, embedded within larger halos. By contrast, for shallow power-spectra with $n<-1$ the density fluctuations at scale $R$ would be governed by the collapse of much larger scales $R_L$ which are just turning non-linear. In the non-linear regime $\sigma(R) \gg 1$ this scale $R_L$ would be much larger than $R$. However, we note that numerical simulations do not show such a transition at $n=-1$. In particular, they are roughly consistent with the stable-clustering ansatz (e.g., \cite{Lac1}, \cite{Col1}) which would be strongly violated in case the pdf would be governed by the saddle-point (\ref{Slamb2}). The reason behind this apparent discrepancy is that the pdf $\cP(\rho_R)$ is not dominated by this spherically symmetric saddle-point. Indeed, this spherical state $\dL$ only governs the pdf if the action remains close to this minimum over a sufficiently large region of phase-space, i.e. for linear density fields which show some slight deviations from this spherical state. For instance, in order to apply the steepest-descent method to rare underdensities in Sect.\ref{Generating function} we had to check that the path-integral (\ref{psib2}) is really dominated by the Gaussian integration around the saddle-point, see the discussion in Sect.\ref{Validity of the steepest-descent method}. We still need to investigate this point for high overdensities. As we explain below, it happens that such a study reveals that the path-integral is not governed by this spherical saddle-point.

\subsection{Virialization processes. Radial-orbit instability}
\label{Virialization processes}

In order to check whether the steepest-descent method around the spherical state obtained in the previous section is valid, we must study the behaviour of the action, hence of the functional $\rho_R[\dL]$, around this point. We may first start by investigating linear perturbation theory. As described below, this will prove sufficient to invalidate the steepest-descent method. This will also provide some useful information about the structure of collapsed halos and virialization processes.

The spherically symmetric saddle-points obtained in Sect.\ref{Spherical overdense saddle-point} exhibit purely radial motions. Therefore, they give rise to non-linear spherical halos with exactly radial orbits. Such orbits are known to be unstable (see for instance \cite{Pal1} and \cite{Pol1} for a presentation of the ``radial-orbit'' instability in some specific cases) hence we can suspect these halos to be unstable. This implies that the action $S[\dL]$ would increase very fast for small non-spherical deviations $\Delta \dL$ around the saddle-point. Then, the density fluctuations at scale $R$ may not be governed by these exactly radial solutions of the dynamics because sufficiently spherical states are too rare. As discussed below this is indeed what occurs in our case. Thus, we study in App.\ref{Radial-orbit instability} the linear perturbation theory around spherical halos with nearly radial orbits. Since the growth rates $\om$ we shall obtain are much larger than the typical frequency $\Omo \sim 1/t_D$ where $t_D$ is the dynamical time (which is of the order of or smaller than the Hubble time $t_H$) the growth of the halo over the time scale $t_H$ is irrelevant. Therefore, we consider static spherical halos with radial orbits. Note that this is rather different from the systems investigated in previous works (e.g., \cite{Pal1}, \cite{Pol1}) where radial orbits only involved a small fraction of the matter content of the halo. In particular, as explained in App.\ref{Radial-orbit instability} while for such systems the authors found slow growth rates $\om \ll \Omo$ as the perturbations develop through a resonance $2:1$, in our case the instability is much more violent because it involves the whole halo and it leads to very high growth rates $\om \gg  \Omo$.

Indeed, considering a halo with nearly radial orbits, or more precisely where the typical angular momentum $\mu$ of the particles is very small, we show in App.\ref{Radial-orbit instability} that non-spherical perturbations are strongly unstable with a growth rate of order:
\beq
\om \sim \Omo \; \sqrt{\frac{L_0}{\mu}} \;\;\; \mbox{if} \;\;\; \mu \ll L_0 ,
\label{omg1}
\eeq
where $\Omo$ is the typical orbital frequency of the particles (see eq.(\ref{Om01})) and $L_0$ is the typical angular momentum of a generic orbit where the transverse and radial velocities are of the same order (see eq.(\ref{limom1})). Besides, the analysis detailed in App.\ref{Radial-orbit instability} is very general. Indeed, it does not rely on the shape of the equilibrium density profile $\rho_0$ nor on the distribution function $f_0$. As a consequence, this radial-orbit instability holds as long as $\mu \ll L_0$. This implies that the halo eventually reaches an equilibrium state where the transverse velocity $\vp$ is of the same order as the radial velocity $v_r$, that is the system becomes roughly isotropic. Moreover, this relaxation is very fast since the growth rate $\om$ diverges for $\mu \rightarrow 0$. Indeed, from eq.(\ref{omg1}) the typical angular momentum $\mu$ of the particles grows as:
\beq
\frac{\d\mu}{\d t} \sim \Omo \; \mu \; e^{\sqrt{L_0/\mu} \; \Omo t} .
\label{dmu1}
\eeq
This merely expresses the fact that the angular momentum of the particles increases with the time-dependent perturbed gravitational potential $\Phi_1(t)$. In eq.(\ref{dmu1}) there may be an additional power-law prefactor however this is irrelevant since the physics is governed by the exponential term which is given by eq.(\ref{omg1}). On a small time-interval $\Delta t \ll t_D$, where $t_D \sim 1/\Omo$ is the dynamical time, we have $t \simeq t_D$ and the growth of the angular momentum is well described by:
\beq
\frac{\d\mu}{\d t} = \Omo \; \mu \; e^{\sqrt{L_0/\mu}} .
\label{dmu2}
\eeq
Then, the time it takes for the angular momentum to grow from $0^+$ up to $\lambda L_0$, with $\lambda\sim 1$, is:
\beq
T(0^+ \rightarrow \lambda L_0) = \frac{1}{\Omo} \int_{0}^{\lambda} \frac{\d y}{y} \; e^{-1/\sqrt{y}} .
\label{dtau1}
\eeq
This integral converges (very fast) for $y \rightarrow 0$, hence it takes a finite time ($T \sim 1/\Omo$) in order to go from $\mu=0^+$ up to $\mu \sim L_0$. This implies that within one dynamical time the typical angular momentum reaches values of order $L_0$, whatever close to exactly radial the system starts from. Note that this is very different from the usual power-law growth of density fluctuations in the expanding universe, where at a finite time $t_0$ the perturbations can be made small enough by starting with a system which is sufficiently close to uniform. By contrast, from eq.(\ref{dtau1}) we see that the system seen after one dynamical time is roughly isotropic ($\vp \sim v_r$), whatever small (but non-zero) the initial transverse velocities are. This implies that the functional $\rho_R[\dL]$ is not continuous at the spherical saddle-point derived in Sect.\ref{Spherical overdense saddle-point}. Indeed, an isotropic velocity distribution provides additional support against the pull from the potential well. This means that the halo is somewhat more extended than the purely spherical radial solution would suggest. This is especially true for power-spectra with $n<-1$ where the radial solution cannot stabilize and leads to a slow adiabatic growth of the density as particles steadily sinks towards the center of the potential well. By contrast, the transverse velocity of the particles stabilizes the density profile and the typical radius of each particle. For instance, \cite{Tey1} find that in the case of spherical gas collapse (i.e. a strongly collisionless fluid with an isotropic pressure) the density profile stabilizes down to $\epsilon > 1/6$ while for $\epsilon <1/6$ the isotropic pressure is insufficient to stabilize the halo which exhibits a density profile of the form $\rho(r,t) \propto t^{(4-24\epsilon)/(18\epsilon)} r^{-1}$, compare with eq.(\ref{rhor1}) and eq.(\ref{rhor2}). We shall come back to this point in Sect.\ref{Virialized halos}.

Therefore, we cannot apply the steepest-descent method around the spherical saddle-point obtained in Sect.\ref{Spherical overdense saddle-point} to the path-integral (\ref{psib2}). Indeed, as explained above, since the functional $\rho_R[\dL]$ is not continuous at this point the spherical dynamics only applies to exactly spherical (hence radial) linear density fields which only form a subset of vanishing measure. Then, the value of the functional $\rho_R[\dL]$ for these spherical states does not govern the path-integral (\ref{psib2}). In simpler words, all realistic density fields show some non-zero deviations from exact spherical symmetry which implies, as we proved in App.\ref{Radial-orbit instability}, that the non-linear objects which form in such an environment are not described by the known solution of the exactly spherical dynamics.

On the other hand, we note that the analysis detailed in App.\ref{Radial-orbit instability} shows that collapsed halos quickly ``virialize'', in the sense that within a dynamical time their velocity distribution becomes roughly isotropic. This is somewhat similar to ``violent relaxation'': starting from an initial state which is very far from thermodynamical equilibrium (the transverse velocity dispersion $\sigma_{\perp}$ is zero) the system undergoes a very fast relaxation phase (over one dynamical time) to reach a new equilibrium state where $\sigma_{\perp}$ is of the order of the radial velocity dispersion $\sigma_r$. This agrees with the results of numerical simulations which show that within one fifth of the virial radius the velocity field is roughly isotropic (e.g., \cite{Tor1}). The virial radius marks the transition between outer infalling shells (which have just experienced their first turn-around) and inner relaxed regions (e.g., \cite{Cole1}).

\subsection{Virialized halos}
\label{Virialized halos}

We have shown in the previous section that the minimum of the action $S[\dL]$ at the spherical saddle-point obtained in Sect.\ref{Spherical overdense saddle-point} does not govern the path-integral (\ref{psib2}). Hence it cannot be used to derive the high-density tail of the pdf $\cP(\rho_R)$. Thus, in order to estimate $\cP(\rho_R)$ we need to know the behaviour of the functional $\rho_R[\dL]$ for non-spherical states. In particular, we are interested in the action $S_c[\dL]$ and the functional $\rho_{R,c}[\dL]$ defined by continuation at spherical points $\dL$. That is, $\rho_{R,c}[\dL]$ at such a spherical point is defined by the value obtained for a state with an infinitesimal deviation $\Delta \dL$ from spherical symmetry in the limit $\Delta \dL \rightarrow 0$. Then, we can still expect this action $S_c[\dL]$ to show a minimum (or a saddle-point) at a spherical state. Indeed, as the deviation from spherical symmetry increases the non-linear halo should be more extended with an irregular boundary so that a larger fraction of the matter lies outside of the radius $R$. Then, it may be justified to apply the steepest-descent method around this point. In case one cannot define a continuous action $S_c[\dL]$ (i.e. the limit $\rho_{R}[\dL]$ depends on the direction along which one approaches exact spherical symmetry) we can still expect it can be approximated with a reasonable accuracy by a smooth functional. This simply means that high-density fluctuations probed by $\cP(\rho_R)$ should be dominated by such nearly spherical linear density states which yield this overdensity $\rho_R$ within the radius $R$. However, in order to be valid this picture requires that these roughly isotropic spherical halos are stable (or only weakly unstable). This ensures that a sufficiently large region of phase-space (i.e. a subset of initial conditions of non-zero measure) is governed by this peculiar dynamics, so that it is indeed relevant to the formation of massive collapsed halos.

As we showed in Sect.\ref{Virialization processes}, almost spherical non-linear density fluctuations become roughly isotropic within a dynamical time. Therefore, since we are interested here in the high-density tail of the pdf $\cP(\rho_R)$, that is in halos with a dynamical time which is smaller than the Hubble time, we consider collapsed objects which have already relaxed. Thus, we merely need to investigate the stability of such relaxed isotropic halos. As noticed in Sect.\ref{Virialization processes}, the transverse velocity dispersion $\sigma_{\perp}$ provides some additional support against the gravitational attraction hence the orbits should stabilize to a finite radius after a few dynamical times. Therefore, we expect the stabilized behaviour (\ref{rhor1}) to apply down to $\epsilon \rightarrow 0$, that is $n \rightarrow -3$. We shall come back to this point below. 

Note that the validity of the density profiles (\ref{rhor1}) rests on the assumption that although the velocities quickly relax to an isotropic distribution the order of magnitude of the energy of most particles remains unchanged during the transition phase. We did not obtain a rigorous proof of this feature (since the relaxation occurs on a dynamical time the evolution is not adiabatic) but it seems quite reasonable. Indeed, the picture we have in mind is that inner shells have already relaxed when new shells fall in (since the dynamical time associated with each shell scales as the time of first turn-around). Then, the new particles quickly acquire a significant transverse velocity during their first infall so that they ``stabilize'' onto orbits with a characteristic radius which is of order of their turn-around radius. This implies that the density profile of the inner regions is not significantly changed and that the energy of these outer particles remains roughly constant. This is to be expected since there is no other energy scale in the problem (for a nearly power-law initial condition the dynamics is roughly self-similar). This also agrees with numerical studies of ```violent relaxation'' in other contexts (\cite{Kan1}).

\begin{figure}[htb]

\centerline{\epsfxsize=8 cm \epsfysize=4.5 cm \epsfbox{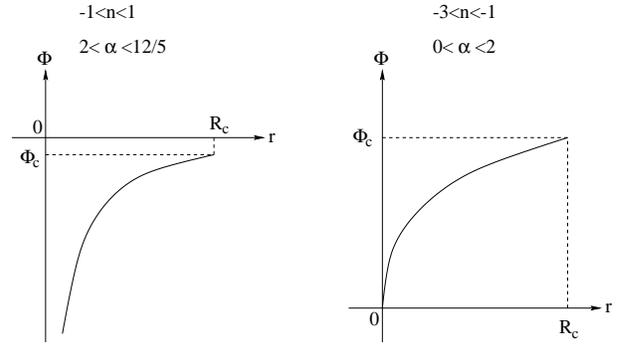}}

\caption{The shape of the gravitational potential $\Phi(r)$ for $0<r<R_c$. For $n>-1$ we have $\Phi(r) \rightarrow -\infty$ in the limit $r\rightarrow 0$ while for $n<-1$ we have $\Phi(r) \rightarrow 0$. The plot on the right shown for the case $0<\alpha<2$ actually corresponds to $1<\alpha<2$. For $0<\alpha<1$ we have the same behaviour (i.e $\Phi(r) \rightarrow 0$ for $r\rightarrow 0$) but the curve looks like $r^2$ rather than $\sqrt{r}$.}
\label{figPhic}

\end{figure}

Thus, we now consider isotropic halos of radius $R_c$ with the density profile:
\beq
\rho(r) = \rho_c \left(\frac{r}{R_c}\right)^{-\alpha} \;\;\; \mbox{with} \;\;\; \alpha = \frac{9\epsilon}{1+3\epsilon} = \frac{3(n+3)}{n+4}
\label{rhoeq1}
\eeq
since from eq.(\ref{spherL1}) and eq.(\ref{profil4}) we have $\epsilon=(n+3)/3$. Note that for power-spectra of cosmological interest which obey $-3<n<1$ we have $0<\alpha<12/5$. Of course, the radius $R_c$ grows with time as new shells turn-around but this is not important as the mass within a given radius is not governed by the outer shells, as we shall check below. Besides, the density profile of the saddle-point is actually flat within $R_L$ but this is irrelevant here since we only want to check whether the overdensity $\rho_R$ is stable with respect to the collapse of the outer shells. The gravitational potential is obtained from eq.(\ref{rhoeq1}) through the Poisson equation which yields:
\beq
\Phi(r) = \Phi_c \left(\frac{r}{R_c}\right)^{2-\alpha} \;\;\; \mbox{with} \;\;\; \Phi_c = \frac{4\pi\cG\rho_c R_c^2}{(2-\alpha)(3-\alpha)} 
\label{Phieq1}
\eeq
for $r<R_c$. Note that with this normalization the value of the gravitational potential at infinity is some constant $\Phi_{\infty}$ which is not zero, but it is irrelevant for our purposes. The shape of $\Phi(r)$ is displayed in Fig.\ref{figPhic}. Thus, we see that the cases $n<-1$ and $n>-1$ are again rather different, as could be expected.

Next, we need the distribution function $f(\br,\bv)$ which describes these collapsed halos. Indeed, our goal is to show that such distributions exist and that they are stable. Since the velocity distribution is isotropic we know that the distribution function only depends on the energy $E$ of the particles (e.g., \cite{Bin1}): $f=f(E)$. Then, the function $f(E)$ can be obtained from the density profile $\rho(r)$ as follows, see \cite{Bin1}. The energy $E$ of the particles decreases for smaller $r$ as they are more strongly bound to the potential well. Moreover, for a halo of size $R_c$ the energy of the particles which reach the radius $R_c$ is $E=\Phi(R_c)=\Phi_c$. Indeed, if they had a larger energy they would have a non-zero velocity at this point which would imply that some particles can move outside of $R_c$, since the velocity distribution is isotropic. Then, we define the new variables:
\beq
\psi(r) \equiv \Phi_c - \Phi(r) \geq 0 , \;\;\; \cE \equiv \Phi_c - E = \psi(r) - \frac{v^2}{2} \geq 0 .
\label{Epsi1}
\eeq
Thus, we write the distribution function $f(E)$ as $f(\cE)$, and we have $f(\cE)=0$ for $\cE<0$. Besides, the density $\rho(r)$ is obtained from $f(\br,\bv)$ by:
\beq
\rho(r) = \int \d^3 v \; f(\br,\bv) = 4\pi \int_0^{\psi(r)} \d\cE \; \sqrt{2(\psi-\cE)} \; f(\cE) .
\label{rhoeq2}
\eeq
Then, since both $\rho(r)$ and $\psi(r)$ are monotonic functions of $r$ we can regard $\rho$ as a function of $\psi$. Next, differentiating both sides of eq.(\ref{rhoeq2}) with respect to $\psi$ we obtain an Abel integral equation which can be inverted as (\cite{Bin1}):
\beq
f(\cE) = \frac{1}{\sqrt{8}\pi^2} \; \frac{\d}{\d\cE} \int_0^{\cE} \d\psi \; \frac{1}{\sqrt{\cE-\psi}} \; \frac{\d\rho}{\d\psi} .
\label{feq1}
\eeq
From eq.(\ref{rhoeq1}) and eq.(\ref{Phieq1}) we obtain:
\beq
\frac{\d\rho}{\d\psi} = \frac{\alpha}{2-\alpha} \; \frac{\rho_c}{\Phi_c} \; \left( 1 - \frac{\psi}{\Phi_c}\right)^{-2/(2-\alpha)}
\label{feq2}
\eeq
which yields:
\beqa
f(\cE) & = & \frac{1}{\sqrt{8}\pi^2} \; \frac{\alpha}{2-\alpha} \; \frac{\rho_c}{\Phi_c} \nonumber \\ & & \times \; \frac{\d}{\d\cE} \int_0^{\cE} \d\psi \; \frac{1}{\sqrt{\cE-\psi}} \; \left( 1 - \frac{\psi}{\Phi_c}\right)^{-2/(2-\alpha)} .
\label{feq3}
\eeqa
We are only interested in the behaviour of $f(\cE)$ which is relevant for small radii $r \ll R_c$ where the halo is fully relaxed. Indeed, the cutoff of the density profile at the boundary $R_c$ is not exact as we should include the transition towards the outer shells which are still falling in. However, here we only investigate the stability of the inner regions which are described by power-law behaviours. Let us note $I_{\cE}$ the integral over $\psi$ which appears in the r.h.s. in eq.(\ref{feq3}). In order to evaluate $I_{\cE}$ we must separate the cases $n>-1$ and $n<-1$ which exhibit different behaviours.

First, in the case $n>-1$ the energy $E$ of the particles which orbit in the inner regions of the halo goes to $-\infty$ as $r\rightarrow 0$, following the divergence of the gravitational potential ($\Phi \rightarrow -\infty$). Therefore, we are interested in the behaviour of $I_{\cE}$ for $\cE \rightarrow +\infty$, see eq.(\ref{Epsi1}). Making the change of variable $\psi = \cE t$ we obtain for $\cE \rightarrow +\infty$:
\beq
I_{\cE} \simeq \cE^{1/2} \left(\frac{\cE}{|\Phi_c|}\right)^{2/(\alpha-2)} \int_0^1 \frac{\d t}{\sqrt{1-t}} \; t^{2/(\alpha-2)}
\label{IE1}
\eeq
where the integral over $t$ is simply the Euler Beta function $B(1/2,\alpha/(\alpha-2))$, see \cite{Grad1}. Thus, we get for large $\cE$:
\beq
n>-1: \;\; f(\cE) \simeq N_{\alpha} \frac{\rho_c}{|\Phi_c|^{3/2}} \left(\frac{\cE}{|\Phi_c|}\right)^{(6-\alpha)/[2(\alpha-2)]}
\label{feq4}
\eeq
with:
\beq
N_{\alpha} = \frac{1}{\sqrt{8}\pi^2} \; \frac{\alpha(\alpha+2)}{2(\alpha-2)^2} \; B\left( \frac{1}{2},\frac{\alpha}{\alpha-2} \right) .
\eeq
Going back to the variable $E$ we can write eq.(\ref{feq4}) as:
\beqa
\lefteqn{-1<n<1, \; 2<\alpha<12/5, \; E \ll -|\Phi_c| :} \nonumber \\ & & f(E) \simeq N_{\alpha} \frac{\rho_c}{|\Phi_c|^{3/2}} \left(\frac{-E}{|\Phi_c|}\right)^{(6-\alpha)/[2(\alpha-2)]} .
\label{feq5}
\eeqa
This gives the behaviour of the distribution function $f(E)$ for large negative $E$, which corresponds to the particles orbiting well inside the halo (i.e. $r\ll R_c$).

Second, in the case $n<-1$ the energy $E$ and the gravitational potential $\Phi$ of the particles which orbit close to the center of the halo vanish. Thus, we are interested in the behaviour of $I_{\cE}$ in the limit $\cE \rightarrow \Phi_c^-$. Using the changes of variables $\psi=\cE t$ and $\cE=\Phi_c x$ we obtain for $x \rightarrow 1^-$:
\beq
I_{\cE} \simeq \Phi_c^{1/2} \int_0^1 \frac{\d t}{\sqrt{1-t}} \; (1-x t)^{-2/(2-\alpha)} .
\label{IE2}
\eeq
The integral in the r.h.s. can be written in terms of Gauss' Hypergeometric function $_2F_1$ (see \cite{Grad1}, \S 3.197.3). Then, using the asymptotic behaviour of the Hypergeometric function for $x \rightarrow 1^-$ we get:
\beq
n<-1: \;\; f(\cE) \simeq M_{\alpha} \frac{\rho_c}{\Phi_c^{3/2}} \left(1 - \frac{\cE}{\Phi_c}\right)^{-(6-\alpha)/[2(2-\alpha)]}
\label{feq6}
\eeq
with:
\beq
M_{\alpha} = \frac{1}{\sqrt{8}\pi^2} \; \frac{\alpha(\alpha+2)}{2(2-\alpha)^2} \; B\left( \frac{1}{2},\frac{2+\alpha}{2(2-\alpha)} \right) .
\eeq
This yields for the distribution function $f(E)$:
\beqa
\lefteqn{-3<n<-1, \; 0<\alpha<2, \; 0 < E \ll \Phi_c :} \nonumber \\ & & f(E) \simeq M_{\alpha} \frac{\rho_c}{\Phi_c^{3/2}} \left(\frac{E}{\Phi_c}\right)^{-(6-\alpha)/[2(2-\alpha)]} .
\label{feq7}
\eeqa
This describes again the orbits which are located within the inner regions of the halo. Note that for $n>-1$ the energy spans the range $]-\infty,-|\Phi_c|]$ while for $n<-1$ it is restricted to $[0,\Phi_c]$.

Firstly, we can see that in both cases the distribution functions $f(E)$ we obtained are positive since $N_{\alpha}>0$ and $M_{\alpha}>0$. Hence they correspond to realistic and physical distributions. Secondly, we can check that the density $\rho(r)$ at radius $r$ is dominated by the ``local'' particles with an orbit of size $\sim r$ and not by the outer shells of radius $\sim R_c$. Thirdly, we can see from eq.(\ref{feq5}) and eq.(\ref{feq7}) that in both cases the distribution function $f$ is a decreasing function of the energy, that is we have $\d f/\d E<0$, over the range where $f$ is not zero. Then, we know that the property $\d f/\d E<0$ ensures that the isotropic equilibrium distribution $f(E)$ is stable, see for instance \cite{Bin1} or \cite{Kan2}. Therefore, we can conclude that the density profiles (\ref{rhoeq1}) are stable for $-3<n<1$. Thus, after their first turn-around the collapsing shells quickly acquire a transverse velocity dispersion $\sigma_{\perp}$ of the same order as the radial velocity dispersion and the particles stabilize to a finite radius. This leads to the density profiles (\ref{rhoeq1}) which agree with stable-clustering and the collapse of new outer shells no longer leads to a slow ``compression'' of the inner mass shells, even in the range $-3<n<-1$. Here it is interesting to note that \cite{Tey1} found that the density profiles (\ref{rhoeq1}) only hold for $\alpha>1$ in the case of gas collapse. Therefore, collisionless halos are more stable than their hydrodynamical (i.e. gaseous) counterparts. This is actually a rather general fact (e.g., \cite{Bin1}). Note that the physical processes at work are quite different. Indeed, for a gaseous system particles are supported against gravity by their numerous collisions with neighbouring particles (which gives rise to the isotropic pressure) while in a collisionless halo each particle is ``stabilized'' within the potential well by its own kinetic energy.

\subsection{High density tail of the pdf $\cP(\rho_R)$}
\label{High density tail of the pdf}

As explained in the previous sections, the high-density tail of the pdf $\cP(\rho_R)$ should be governed by almost spherical halos which have relaxed to a roughly isotropic equilibrium distribution function. The density profile is stabilized by the transverse velocity dispersion so that stable clustering holds. Moreover, these equilibrium states are stable (at linear order) with respect to small perturbations. Therefore, we can now apply to the whole range $-3<n<1$ the considerations used in Sect.\ref{Spherical overdense saddle-point} for $n>-1$. More precisely, we approximate the functional $\rho_R[\delta_{L,R''}]$ by eq.(\ref{F1}) where the Lagrangian scale $R_L$ is associated with the particles with a turn-around radius equal to $R$. The function $\cFr(\dL)$ now describes the continuous limit of almost exactly spherical collapse. In other words, it is not given by the exact spherical dynamics with radial trajectories: it is defined as the non-linear overdensity $\rho_R$ obtained in the limit of infinitesimal deviations from spherical symmetry. Thus, it corresponds to the virialized distribution functions $f(E)$ obtained in Sect.\ref{Virialized halos}. 

As recalled in Sect.\ref{Spherical saddle-point}, a relation of the form (\ref{F1}) leads to a spherical saddle-point given by eq.(\ref{col1}) and eq.(\ref{col2}). Provided the steepest-descent approximation is justified in the limit we consider here (i.e. $\rho_R \rightarrow \infty$) this yields for the generating function $\psib(y)$ the expression (\ref{psib3}), where eq.(\ref{M1}) through eq.(\ref{G3}) must be used with the relevant function $\cFr(\dL)$. We refer again the reader to \cite{paper2} for a detailed derivation of these points. Thus, we now need to write an explicit expression for the function $\cFr(\dL)$. Since after virialization we assume the density profile to remain stable the overdensity $\rho_R$ grows as $\rho_R \propto \rhob^{-1} \propto a^3$ while $\dL(M) \propto a$ (here we consider a critical density universe). Hence we get $\rho_R \propto \dL^3$ and we write:
\beq
\cFr(\dL) = (1+\Dc) \left( \frac{\dL}{\dc} \right)^3 .
\label{Fv1}
\eeq
Note that this is very similar to the usual Press-Schechter prescription (\cite{PS}). In particular, the normalization parameters $\Dc$ and $\dc$ should be close to the usual values $\dc \simeq 1.69$ and $\Dc \simeq 177$. However, these latter values are only approximate. In order to obtain the right normalization one should run a numerical simulation in order to study the collapse of a density profile of the form (\ref{col2}), to which is added a small deviation from spherical symmetry. In other words, one must obtain the behaviour of the saddle-point of the continuous action $S_c[\dL]$. However, such a numerical study is beyond the scope of this paper. Note that we can expect a small dependence on the slope $n$ of the power-spectrum of the normalization $(1+\Dc)/\dc^3$. From eq.(\ref{G3}) we obtain the characteristic function $\cGr(\tau)$ as:
\beq
\tau \ll -1 : \;\; \cGr(\tau) = \left( \frac{1+\Dc}{\dc^3} \right)^{2/(n+5)} \; (-\tau)^{6/(n+5)} . 
\label{Gv1}
\eeq
Note that large overdensities $\rho_R$ are associated with $\tau \rightarrow -\infty$, see \cite{paper2}. Then, from eq.(\ref{Sy1}) we have the asymptotic behaviour for $\rho_R \rightarrow +\infty$:
\beq
\rho_R = \cGr(\tau) \simeq \left( \frac{1+\Dc}{\dc^3} \right)^{2/(n+2)} \; \left( \frac{-6\;y}{n+5}\right)^{3/(n+2)} .
\label{rhoyd1}
\eeq
Note that for $n>-2$ large positive $\rho_R$ are associated with large negative $y$. This corresponds to a pdf $\cP(\rho_R)$ which decreases faster than a pure exponential at large densities. By contrast, for $n<-2$ high densities are associated with $y \rightarrow 0^-$, which gives rise to a cutoff which is smoother than an exponential. This is discussed at length in \cite{paper2}, where we come across the same transition at $n=0$ in the quasi-linear regime.

As for the case of underdensities described in Sect.\ref{Generating function} we now need to check whether the steepest-descent method is justified in the limit of high densities. This is again a difficult point since we do not know the exact form of the functional $\rho_R[\dL]$. However, we can apply again the discussion developed in Sect.\ref{Validity of the steepest-descent method} about the ordinary integral (\ref{psi11}) since we have similar power-law behaviours. Then, the generating function $\psib(y)$ is still given by the expression (\ref{psib4}) and as in Sect.\ref{Asymptotic form of the generating function} we use the approximation (\ref{psib5}) from the analogy with the one-dimensional case, taking into account the expansion of the universe. Note that the factor $1/\rho_R$ is now associated with ``sharp'' directions (the translations). Then, we obtain again the expression (\ref{P5}) for the pdf $\cP(\rho_R)$. Using the power-law behaviour (\ref{Gv1}) this yields:
\beqa
\cP(\rho_R) & \simeq & \frac{1}{\sqrt{2\pi}\sigma} \; \frac{n+5}{6} \; \left( \frac{1+\Dc}{\dc^3} \right)^{-1/3} \; \rho_R^{-(7-n)/6} \nonumber \\ & & \times \; e^{-(\frac{1+\Dc}{\dc^3})^{-2/3} \rho_R^{(n+5)/3} /(2\sigma^2)} .
\label{Pov1}
\eeqa
As noticed above, for $n<-2$ the saddle-point is actually a local maximum of the action so that the steepest-descent method requires some care. This is similar to what occurs for $n<0$ in the quasi-linear regime. Thus we refer the reader to App.A in \cite{paper2} for a detailed discussion of the way to handle such cases. However, eq.(\ref{Pov1}) remains valid for all $n$.

Of course, as for the case of underdensities discussed in Sect.\ref{Spherical model} the expression (\ref{Pov1}) agrees with the simple spherical model (\ref{spher1}), where the spherical dynamics used to relate the linear and non-linear density contrasts is given by eq.(\ref{Fv1}). Thus, we recover the pdf obtained in \cite{Val1} and \cite{Val2} from the spherical model coupled with the stable-clustering ansatz. As discussed in Sect.\ref{Asymptotic form of the generating function}, the prefactor which appears in eq.(\ref{Pov1}) is not exact because we used a simple approximation for the determinant $\cD$ which arises from the Gaussian integration of the path-integral around the spherical saddle-point. In particular, we can expect the exact calculation to give a numerical multiplicative factor of order unity. However, we note that the same approach used in the quasi-linear regime provides very good results as compared to numerical simulations (e.g., \cite{Val2}, \cite{paper2}). Therefore, the correction might be quite small (but this is a different regime hence we cannot draw definite conclusions).

The pdf $\cP(\rho_R)$ shows a cutoff at the characteristic density $\rho_c$ such that $\tau_c \sim -\sigma$. In the quasi-linear regime this corresponds to $\delta_R \sim \sigma$ while in the non-linear regime where we can use eq.(\ref{Gv1}) this yields $\rho_R \sim \sigma^{6/(n+5)}$. Therefore, we can write the cutoff $\rho_c$ as:
\beq
\rho_c = \mbox{Min} \left[ 1 + \sigma(R) , \sigma(R)^{6/(n+5)} \right] .
\label{rhoc1}
\eeq
which is valid for all values of $\sigma$. Indeed, we recall that we only used the limit of rare overdensities and the value of $\sigma$ is actually irrelevant, so that our results apply from the quasi-linear regime ($\sigma \ll 1$) up to the highly non-linear regime ($\sigma \gg 1$). The influence of $\sigma(R)$ only appears in the cutoff $\rho_c$. As we probe deeper into the non-linear regime the characteristic overdensity $\rho_c$ of virialized halos is pushed to higher values. In the highly non-linear regime, eq.(\ref{rhoc1}) gives:
\beq
\sigma(R) \gg 1 : \; \rho_c = \sigma(R)^{\frac{6}{n+5}} = \left(\frac{R}{R_0}\right)^{-\frac{3(n+3)}{n+5}} \sim \xib(R)
\label{rhoc2}
\eeq
where we noted $R_0$ the non-linear scale (defined by $\sigma(R_0)=1$) and we used eq.(\ref{gam1}) for the last term. Here, we must point out that we only derived eq.(\ref{Pov1}) in the limit of rare overdensities. That is, it only applies to large densities with $\rho_R \gg \rho_c$. The behaviour of the pdf $\cP(\rho_R)$ at lower densities cannot be obtained by the steepest-descent method detailed in this article. Thus, for intermediate densities $\rho_v < \rho_R <\rho_c$ the path-integral (\ref{psib2}) is no longer dominated by one saddle-point. Then, one must devise another approach which takes into account the dynamics of a wide range of initial conditions $\dL(\bx)$. Going back to the interpretation of the results (\ref{P6}) and (\ref{Pov1}) in terms of the simple spherical model this is also quite clear. Indeed, these intermediate densities correspond to relatively low densities (i.e. smaller than the typical density $\rho_c\sim\xib$ associated with collapsed halos of size $R$) where virialization processes occured lately. Then, the dynamics of these objects must have been strongly influenced by the gravitational interaction (e.g., mergings, tidal effects) with neighbouring halos which are typically more massive.

\subsection{Aging processes ?}
\label{Aging processes}

The very dense halos with $\rho_R \gg \rho_c$ are sufficiently rare and massive not to be affected by the interaction with neighbours as they form. However, we note that the present context shows some important differences with the case of rare underdensities studied in Sect.\ref{Rare underdensities}. Indeed, in that former case the underdensities which govern the low-density tail of the pdf consist of rare regions which are still expanding in the universe and the dynamics is fully determined by the spherical model. Moreover, outer regions have no strong influence on the behaviour of the underdensity. By contrast, in the present case virialized halos remain stable after their collapse. However, this may only be a zeroth order approximation. Indeed, as larger scales turn non-linear and collapse the halo of size $R$ which we study becomes embedded within more massive and extended structures. Then, as time goes on we can expect repeated interactions with neighbouring halos and tidal effects (e.g., mergers, variations of the large scale gravitational potential) to have a cumulative impact onto this object which will gradually be distorted or even disrupted or merged within a larger halo. Note that these ``aging processes'' should be less efficient for more massive objects. Therefore, a possible scenario would be that the high-density tail we obtained in eq.(\ref{Pov1}) only applies to densities which are larger than a time-dependent threshold $\rho_t$ which increases faster than $\rho_c$ with time. Thus, we would have $\rho_t \sim \rho_c$ for $\xib \sim 1$, when larger scales have not collapsed yet, and $\rho_t/\rho_c \rightarrow \infty$ for $\xib \rightarrow \infty$, when tidal effects (in a broad sense) have had plenty of time to influence the density profile of the halo. 

Let us briefly describe how such a mechanism might show up in the path-integral formalism we used in the previous sections. As seen from eq.(\ref{psib2}) and eq.(\ref{S1}), the term $(\dL.\DL^{-1}.\dL)$ in the exponent of the path-integral leads to a characteristic cutoff $|\delta_{L,R'}| \sim \sigma(R')$ for the non-spherical perturbations at scale $R'$ of the density fields $\dL$ around the saddle-point which provide a significant contribution to $\psib(y)$. These deviations are much smaller than the density $\delta_{L,R'}$ of the saddle-point, in the limit of rare events, for $R' \sim R_L$. Therefore, they have no strong impact on the collapse of the halo. However, on much larger scales they become non-negligible since the density profile of the saddle-point declines much faster at large scales, as $\delta_{L,R'} \propto \sigma(R')^2$ as seen in eq.(\ref{profil4}). Then, when these scales turn non-linear (i.e. much after the halo of Eulerian size $R$ has collapsed and virialized) these important deviations from spherical symmetry may play a key role. Indeed, since the collapse is no longer roughly self-similar but proceeds in a rather irregular manner, because of the break of spherical symmetry, one expects the formation of an irregular distribution of clumps. Then, this could lead to the tidal effects or merging processes described above. Indeed, the regular and adiabatic collapse of outer shells (adiabatic with respect to the inner shells) is replaced by an irregular evolution which may proceed through sudden non-linear changes in the large scale gravitational potential. This means that the region of phase-space where the action $S[\dL]$ is close to its value at the saddle-point becomes narrower as time goes on: one needs to restrict to linear states $\dL$ which are increasingly spherically symmetric. This implies a change in the normalization of the generating function $\psib(y)$ in this regime. Mathematically, this arises from an increase of the determinant $\cD$ which appears in eq.(\ref{psib4}) (transposed to the regime of high overdensities) from the Gaussian integration around the saddle-point. In other words, the action $S[\dL]$ exhibits an increasingly fast variation with $\dL$ as time goes on.

The effect of these processes on the high-density tail of the pdf $\cP(\rho_R)$ is now clear. The exponential cutoff obtained in eq.(\ref{Pov1}) remains valid (at least in a first step) but the prefactor steadily decreases as time goes on, following the decline of the phase-space volume (i.e. the number of states $\dL$) which leads to this non-linear overdensity $\rho_R$. Note however that this decline could be hidden if these tidal effects also build deeper minima of the action. These effects should be stronger for lower overdensities $\rho_R$. Finally, after the pdf $\cP(\rho_R)$ at such a point has undergone a significant evolution we can expect that it enters a new regime (possibly non-stationary) where it is governed by these ``tidal'' processes. It is clear that this regime cannot be obtained by a steepest-descent method similar to the approach developed in this paper. We must also note that these arguments, although quite reasonable, are still hypothetical at this stage for want of a rigorous theoretical derivation. However, this point requires new theoretical tools and a detailed analysis which are beyond the scope of this article.

\subsection{Comparison with the non-linear scaling model}
\label{Comparison with the non-linear scaling model}

Here we briefly compare our result (\ref{Pov1}) with the non-linear scaling model recalled in Sect.\ref{Non-linear scaling model}. Since eq.(\ref{Pov1}) is consistent with the stable-clustering ansatz -as it neglects the possible ``aging processes'' discussed in Sect.\ref{Aging processes}- it is clear that it is also consistent with the scaling laws (\ref{scal1}). Let us recall briefly the main properties of the non-linear scaling model (\ref{scal1}) with respect to the high-density tail of the pdf $\cP(\rho_R)$. As shown in \cite{Bal1}, the scaling laws (\ref{scal1}) imply that for large overdensities $\rho_R \gg \rhot_v$ the pdf exhibits the scaling:
\beq
\rho_R \gg \rhot_v : \;\; \Pt(\rho_R) = \frac{1}{\xib^{\;2}} \; h(x) \;\;\; \mbox{with} \;\;\; x \equiv \frac{\rho_R}{\xib} ,
\label{hx1}
\eeq
where we defined the scaling function $h(x)$ by:
\beq
h(x) \equiv - \inta \frac{\d y}{2\pi i} \; e^{x y} \; \phit(y) ,
\label{hx2}
\eeq
and the function $\phit(y)$ was introduced in eq.(\ref{phit1}). As shown in \cite{Val1}, the expression (\ref{Pov1}) can actually be recast in the form (\ref{hx1}). Indeed, for a power-law linear power-spectrum $P(k)\propto k^n$ the non-linear correlation $\xib(R,t)$ can be written as a function of $\sigma(R,t)$. From eq.(\ref{gam1}) we obtain $\xib \propto \sigma^{6/(n+5)}$. Therefore, we write:
\beq
\sigma \gg 1 : \;\; \xib(R) = \left( \frac{1+\Delta_{\xi}}{\dc^3} \right)^{2/(n+5)} \; \sigma(R)^{6/(n+5)}
\label{Xisig1}
\eeq
where both correlations $\xib$ and $\sigma$ are taken at the same Eulerian scale $R$. We wrote the normalization factor in eq.(\ref{Xisig1}) as in eq.(\ref{Gv1}) and $1+\Delta_{\xi}$ is an unknown parameter which may be taken from numerical simulations. It corresponds to an ``effective'' non-linear density contrast associated with the linear contrast $\dc$. This agrees with the scaling ansatz introduced by \cite{Ham1} to relate the non-linear correlation at scale $R$ to the linear variance $\sigma(R_L)$ at the Lagrangian scale $R_L$. In particular, as noticed in \cite{Val1} (App.E) the relation $\sigma(R_L) \leftrightarrow \xib(R)$ is well described by the spherical collapse model. Thus, eq.(\ref{Xisig1}) actually means that we have $\xib(R) = \cF_{\xi}[\sigma(R_L)]$ where $\cF_{\xi}$ is of the form (\ref{Fv1}) where $\Dc$ is replaced by $\Delta_{\xi}$. Note that we should have $\Delta_{\xi} \la \Dc$ since density fields with important deviations from spherical symmetry should reach a smaller non-linear density contrast $\Delta$ than the value $\Dc$ obtained for the (almost) spherical saddle-point and $\Delta_{\xi}$ corresponds to an average over all possible density fields. Substituting eq.(\ref{Xisig1}) into eq.(\ref{Pov1}) we obtain the scaling law (\ref{hx1}) with:
\beq
h_s(x) \equiv \frac{1}{\sqrt{2\pi}} \; \frac{n+5}{6 \lambda} \; x^{-(7-n)/6} \; e^{- x^{(n+5)/3} /(2\lambda^2)}
\label{hs1}
\eeq
where the subscript ``s'' refers to the spherical model, and we defined the parameter $\lambda$ by:
\beq
\lambda \equiv \left( \frac{1+\Dc}{1+\Delta_{\xi}} \right)^{1/3}  \ga 1 .
\label{lamb1}
\eeq
The parameter $\lambda$ is expected to show some dependence on $n$ since both $\Dc$ and $\Delta_{\xi}$ depend on $n$. This dependence for $\Delta_{\xi}$ was indeed checked in numerical simulations, see for instance \cite{Jain1}.

\begin{figure}

\centerline{\epsfxsize=8 cm \epsfysize=5 cm \epsfbox{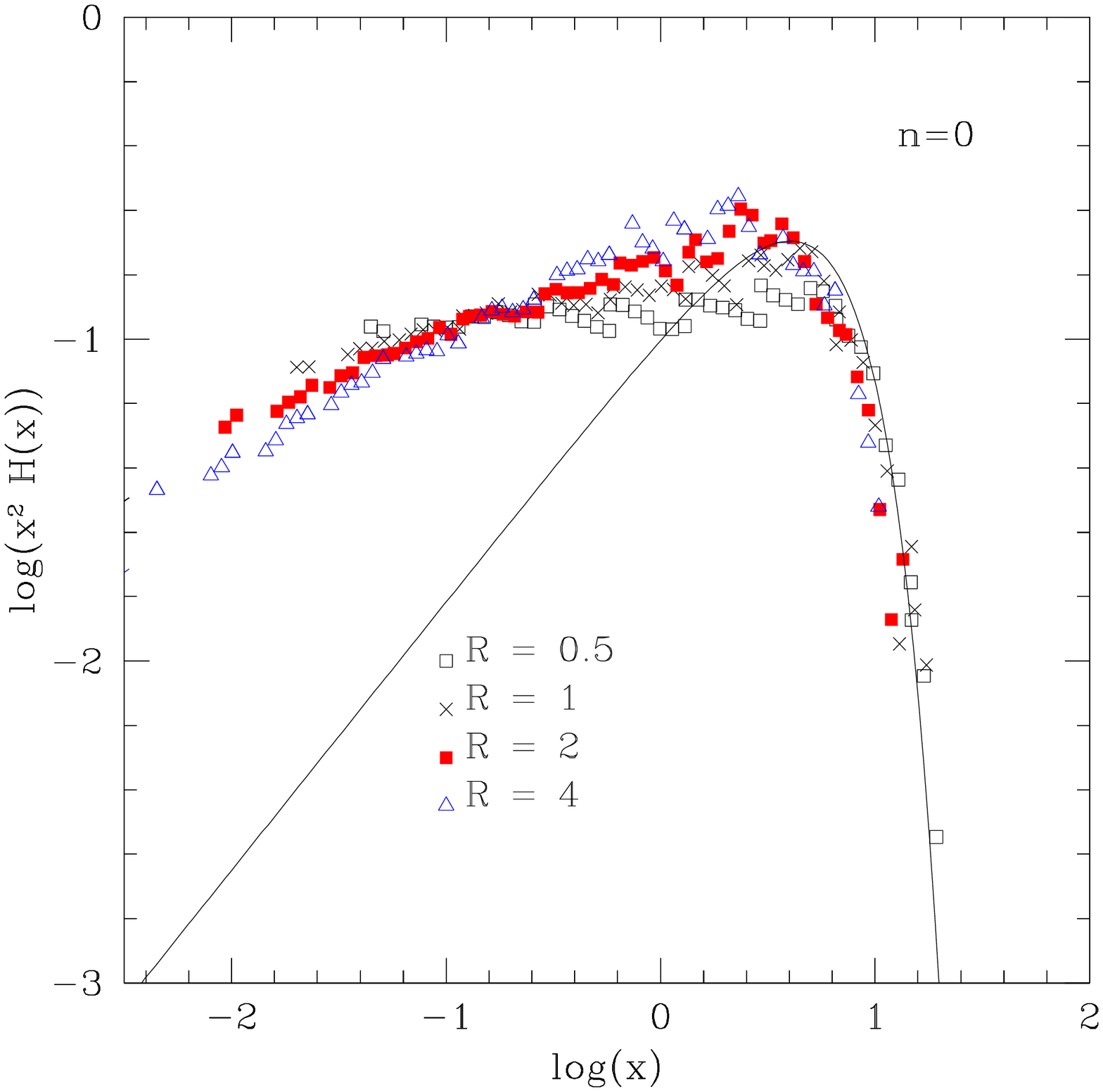}}
\centerline{\epsfxsize=8 cm \epsfysize=5 cm \epsfbox{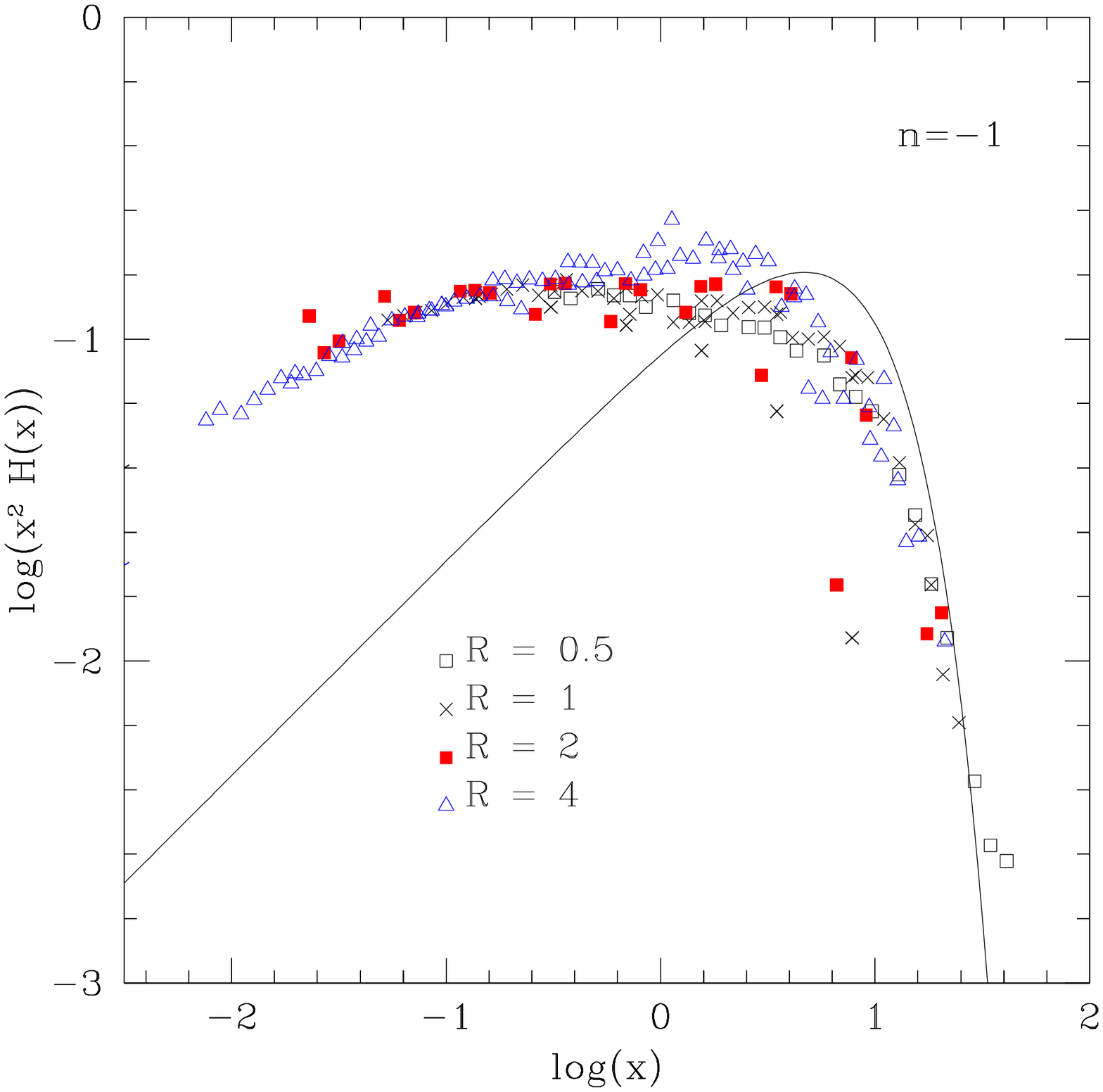}}
\centerline{\epsfxsize=8 cm \epsfysize=5 cm \epsfbox{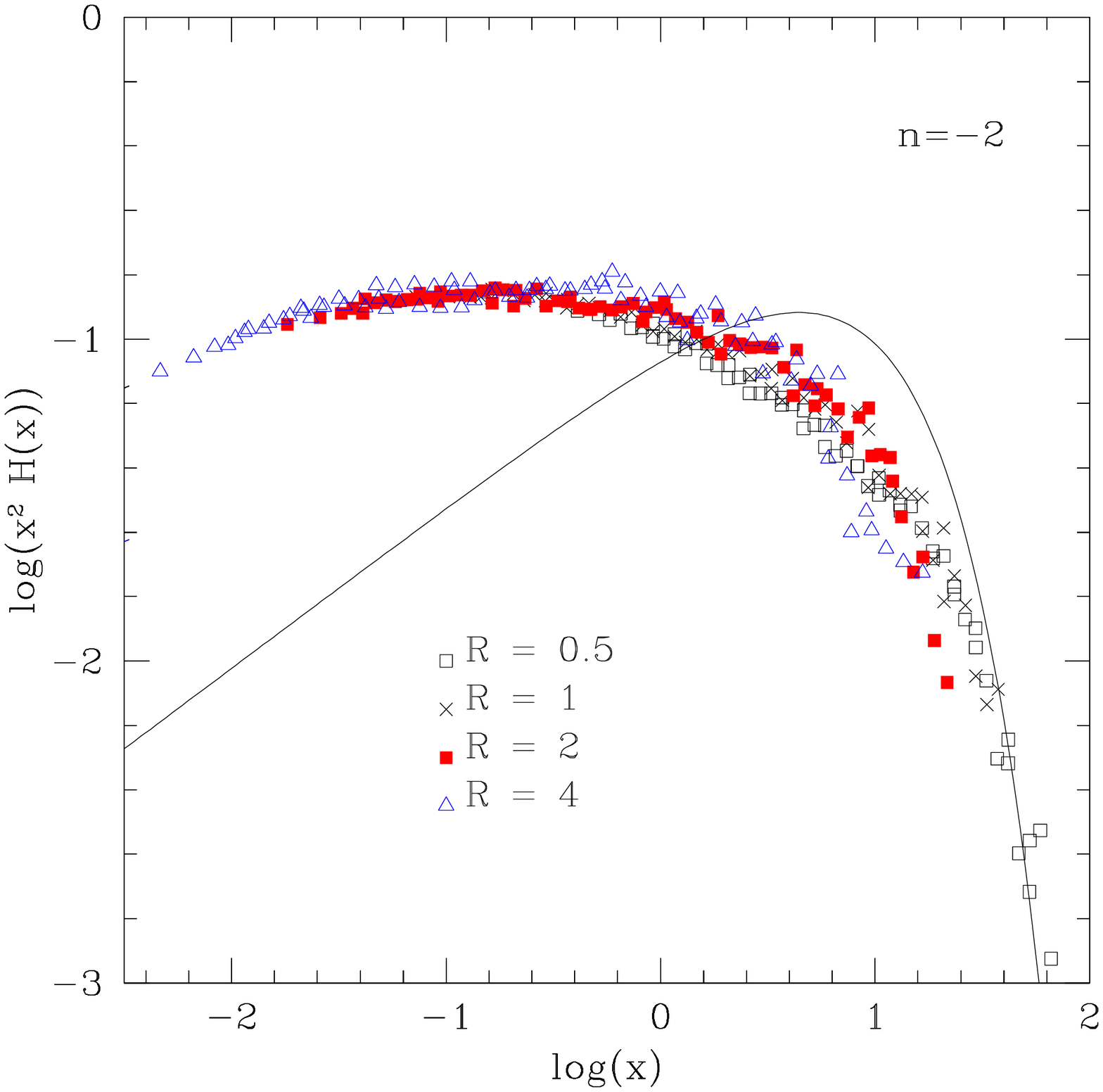}}

\caption{The mass functions of halos defined by various comoving radii $R$, obtained from a ``modified spherical overdensity'' algorithm. The data points are taken from the numerical simulations described in \cite{Lac1}. Different symbols correspond to different values of $R$. Note that for each radius we display the results obtained at several times which correspond to various values of $\xib$. The solid curve is the scaling function $h_s(x)$. We use the values $\lambda=3.2, 2.8$ and $2.1$ for $n=0, -1$ and $-2$.}
\label{figmfR}

\end{figure}

As we pointed out in Sect.\ref{High density tail of the pdf} the expression (\ref{Pov1}) only applies to rare overdensities $\rho_R \gg \rho_c$. In terms of the scaling variable $x$ introduced in eq.(\ref{hx1}) this means that the function $h_s(x)$ written in eq.(\ref{hs1}) only applies to $x \gg 1$. Unfortunately, this makes the comparison with numerical simulations rather difficult since one needs to probe the far high-density tail of the pdf $\cP(\rho_R)$. Indeed, the points measured in current numerical simulations from counts-in-cells statistics do not go much beyond $x \sim 10$, see for instance Fig.4 and Fig.6 in \cite{Lac1}. This is not sufficient to really probe the cutoff of the pdf. However, in \cite{Lac1} we devised a statistics which allows one to go slightly deeper into the high-density tail. Thus, instead of computing the pdf $\cP(\rho_R)$ one can investigate the cumulative mass function $F_R(>M)$ of halos of radius $R$ more massive than $M$. In practice, one uses a ``spherical overdensity algorithm'', looking at particles in order of decreasing density and defining halos as objects of constant size $R$ around the density peaks. Then, the differential mass function $\mu_R(M) \d M/M$ should obey the scaling law:
\beq
\mu_R(M) \frac{\d M}{M} = x^2 H(x) \; \frac{\d x}{x}
\label{HR1}
\eeq
with a scaling function $H(x)$ which is close to $h(x)$, see \cite{Val1} and \cite{Lac1}. In particular, the high-density tails are expected to show the same behaviour (up to a normalization factor of order unity), see \cite{Val1}. This procedure allows one to probe slightly deeper into the high-density tail of $h(x)$ because ``cells'' are directly drawn around high-density peaks. Then, these high-density fluctuations are well accounted for, while for standard counts-in-cells statistics (i.e. the pdf $\cP(\rho_R)$) such high-density peaks are usually split over several cells. Therefore, we compare in Fig.\ref{figmfR} the scaling function $h_s(x)$ obtained in eq.(\ref{hs1}) with the results of numerical simulations taken from \cite{Lac1} (note that this corresponds to the PS prescription without the factor 2). We choose the parameter $\lambda$ so as to get a reasonable fit to the numerical points (see the caption for the relevant values). We can check that we get $\lambda \ga 1$ as it should. The figures show that the numerical simulations are consistent with the exponential tail written in eq.(\ref{hs1}). Note however that we expect the normalization of eq.(\ref{hs1}) to be only approximate because of the approximation (\ref{psib5}) for the determinant $\cD$. Moreover, as noticed above $H(x)$ may differ from $h(x)$ by a factor of order unity, in-between $\gam/3$ and $1$ in simple models (\cite{Val1}).

Note that the data points shown in Fig.\ref{figmfR} are obtained from different comoving radii as well as from the same comoving radius at different times (hence different values of $\xib$ and $\sigma$). Therefore, the fact that all curves superpose shows that the stable-clustering ansatz (\ref{scal1}) is a good approximation in the regime probed by these numerical points. In any case, the point of Fig.\ref{figmfR} is merely to show that eq.(\ref{Pov1}) is consistent with numerical simulations. In fact, it is difficult to see how one could ``beat'' the exponential cutoff given by eq.(\ref{Pov1}). Indeed, it is clear that in the rare-event limit the cutoff of the pdf is directly linked to the initial Gaussian cutoff. Then, in order to change the high-density tail obtained in eq.(\ref{Pov1}) one should rely on a mechanism which would violate eq.(\ref{Fv1}). As explained in Sect.\ref{Aging processes} such a process may indeed exist (through gravitational interactions with outer shells) but it should be negligible if one considers sufficiently large overdensities $\rho_R$ (at a given time). Of course, we can see in Fig.\ref{figmfR} that the pdf (\ref{Pov1}) fails at small densities (i.e. $x \la 1$). As explained in Sect.\ref{High density tail of the pdf} this is quite natural since the steepest-descent method should not apply to this regime. Therefore, in order to derive a prediction for this low-density part of the pdf (which seems to exhibit a power-law behaviour) one needs to build other non-perturbative tools which fully take into account the involved processes of mergings and tidal interactions (with similar or more massive neighbours) which govern the dynamics of these intermediate objects. Here we must note that in \cite{Lac1} we had concluded that eq.(\ref{hs1}) does not agree with the counts-in-cells $\cP(N)$ measured in the simulations. However, we used different values of $\lambda$ and we compared the pdf (\ref{Pov1}) with numerical points over the whole range $\rho_R \gg \rhot_v$. Here we still agree with this conclusion (as noticed above the small$-x$ tail clearly fails) but we note that the high-density cutoff ($x \gg 1$) may be consistent with numerical simulations.

\subsection{Press-Schechter mass function}
\label{Press-Schechter prescription}

Finally, in view of the practical importance of the Press-Schechter (PS) prescription (\cite{PS}) to compute the mass function of just-virialized objects we comment this ``recipe'' in the light of the present work. The PS approach provides a simple estimate of the cumulative mass function $F(>M)$ which gives the fraction of matter embedded within just-virialized halos (i.e. with a non-linear density contrast $\Dc$) of mass larger than $M$. As for the spherical model (\ref{spher1}) it merely approximates $F(>M)$ by the fraction of matter which is enclosed within spherical cells of Lagrangian radius $R_L$ and linear density contrast greater than $\delta_{L,R_L}$, with $\delta_{L,R_L}=\dc$. Therefore, as in eq.(\ref{spher1}) it writes:
\beq
F_{PS}(>M) = \int_{\dc}^{\infty} \d\dL \cP_L(\dL) = \int_{\dc/\sigma(M)}^{\infty} \d\nu \; \cP_L^{(\nu)}(\nu)
\label{PS1}
\eeq
where the subscript ``PS'' refers to the Press-Schechter prescription. This gives the differential mass function $\mu_{PS}(M) \d M/M$ as:
\beqa
\lefteqn{ \mu_{PS}(M) \frac{\d M}{M} = - \frac{\d F_{PS}}{\d M} \; \d M } \nonumber \\ & & = \frac{1}{\sqrt{2\pi}} \; \frac{\dc}{\sigma(M)} \; \left|\frac{\d\ln\sigma}{\d\ln M}\right| \; e^{-\dc^2/(2\sigma(M)^2)} \; \frac{\d M}{M} .
\label{PS2}
\eeqa
Then, in order to obtain a mass function which is correctly normalized to unity it is customary to multiply the expression (\ref{PS2}) by an ad-hoc factor 2.

The quantity we investigated in this article is not the mass function of just-virialized halos but the pdf $\cP(\rho_R)$, that is the statistics of the counts-in-cells. In order to compare our results with the PS mass function we first need a prescription to obtain the mass function from the counts-in-cells. To do so, we can employ the prescription used in the PS approach, which we now apply to the non-linear density field. That is, we simply write:
\beq
F(>M) \simeq \int_{\Dc}^{\infty} \d\delta \; (1+\delta) \cP(\delta) .
\label{FM1}
\eeq
This means that $F(>M)$ is approximately the fraction of matter which is enclosed within spherical cells of radius $R$ and linear density contrast greater than $\Dc$. This is the translation of eq.(\ref{PS1}) in terms of the non-linear density field. Note however that in both cases this is only an approximation since it does not take into account the cloud-in-cloud problem. The relation (\ref{FM1}) is more general and should yield better results since, contrary to eq.(\ref{PS1}), it does not assume that there is a one-to-one relation between the linear and non-linear density contrasts (given by the spherical dynamics for the PS prescription). In particular, as shown in \cite{Val1}, the cloud-in-cloud problem is very severe at small masses when one works with the linear density field as in eq.(\ref{PS1}) so that the low-mass tail of the mass function is actually ill-defined. By contrast, simple models (based on the stable-clustering ansatz recalled in Sect.\ref{Non-linear scaling model}) show that when one works with the actual non-linear density field as in eq.(\ref{FM1}) the cloud-in-cloud problem becomes much less serious since it only yields a correction of order unity for the normalization of the low-mass tail of the mass function, see \cite{Val1}. Of course, the problem with eq.(\ref{FM1}) is that one needs to estimate the non-linear pdf $\cP(\dR)$, which is still an unsolved problem (in this article we only obtained the tails of this pdf).

As noticed in Sect.\ref{High density tail of the pdf}, the expression (\ref{Pov1}) agrees with the simple spherical model (\ref{spher1}) detailed in Sect.\ref{Spherical model}. Therefore, using eq.(\ref{FM1}) we recover the usual PS prescription without the factor 2. Here, we must note that the prefactor which appears in eq.(\ref{Pov1}) is not exact, as discussed above, so that a rigorous calculation of the determinant $\cD$ might yield a factor 2. However, this is rather unlikely and until we have a rigorous estimate of this determinant we can as well keep the normalization obtained in eq.(\ref{Pov1}) or simply fit the normalization to the results of numerical simulations. Note that since the PS mass function deals with ``just-virialized'' halos we do not encounter the ``aging processes'' discussed in Sect.\ref{Aging processes}. Indeed, these halos have just collapsed and they have not had time to suffer from the cumulative effects brought by the interaction with smaller neighbours. Therefore, the ``exponential'' form of the cutoff of the PS mass function should be exact.

We must stress that the expression (\ref{Pov1}) was only derived in the limit of high overdensities, that is for $\rho_R \gg \rho_c$. Thus, as discussed in the previous section, our analysis shows that the exponential cutoff of the PS mass function is correct but the PS approach should not be used for the low-mass (or low-density) tail. In other words, the power-law behaviour which appears in eq.(\ref{Pov1}) for $\rho_R \ll \rho_c$, or for small masses $M \ll M_c$ in the PS mass function, is not justified by the approach developed in this article. In fact, we can actually expect this power-law to be wrong since we know that deviations from spherical symmetry play a key role in this regime. Besides, as we recalled above the cloud-in-cloud problem removes any predictive power to the low-mass tail of the PS mass function (e.g., \cite{Val1}). This implies that the high-mass cutoff of the PS mass function, or the high-density cutoff of the pdf $\cP(\rho_R)$ written in eq.(\ref{Pov1}), should not necessarily be multiplied by a factor 2 in order to obtain a correct normalization of the mass function to unity, since the usual analytical mass function (or pdf) does not extend to $M < M_c$ or $\rho_R < \rho_c$. 

Moreover, the values $\dc=1.69$ and $\Dc=177$ used in the literature are only approximate. We did not derive the exact values in this paper but we explained in Sect.\ref{Virialized halos} how they could be obtained by numerical means. One needs to find the saddle-point $\dL(\bx)$ of the action $S[\dL]$. A simple procedure which should be sufficient for practical purposes is to study through a numerical simulation the dynamics of a spherical linear state of the form (\ref{col2}), to which we add a small perturbation from spherical symmetry. Then, one could measure the function $\cFr(\dL)$ written in eq.(\ref{Fv1}). Of course, we expect to recover values close to the standard density contrasts $\dc=1.69$ and $\Dc=177$ but there should be a small dependence on the slope $n$ of the power-spectrum. Here, we note that recent numerical simulations (e.g., \cite{Gov1}) show that the usual PS mass function (with the factor 2) overestimates the number of small halos and underestimates the number of massive halos. In the light of the work presented in this article, the deviation at small masses (below the cutoff $M_c$) is not surprising since we do not expect the PS mass function to apply to this regime. On the other hand, the exponential tail of the PS mass function should be correct. Then, the discrepancy with the numerical results would be due to the fact that these authors (following the standard practice) do not use the exact values of the density contrasts $(\dc,\Dc)$ nor the correct normalization. Indeed, as we explained above there is no reason to multiply the high-mass tail by a factor 2. Unfortunately, we did not manage to predict accurately this normalization factor. Moreover, there still exists the possibility that an exact calculation of the determinant $\cD$ which appears through the Gaussian integration around the saddle-point modifies the exponent of the power-law prefactor in eq.(\ref{Pov1}) or gives rise to logarithmic corrections.

\section{Conclusion}

Using a non-perturbative method developed in a previous work (\cite{paper2}) we have investigated in this article the tails of the pdf $\cP(\rho_R)$. Our approach is based on a steepest-descent approximation which should yield asymptotically exact results in the limit of rare events. Therefore, it applies to all values of the rms linear density fluctuation $\sigma$, from the quasi-linear up to the highly non-linear regime.

First, we studied the low-density tail of the pdf, that is very rare underdensities. This regime is rather simple since shell-crossing only occurs for power-spectra with a slope $n>-1$, and even in this case shell-crossing only has a limited impact as it merely introduces numerical factors of order unity in the pdf but it should not modify the characteristic exponents which govern the low-density cutoff. This allows us to get a good description of rare underdensities and to derive the shape of the low-density tail of the pdf. Apart for the power-law prefactor (which we did not derive in a rigorous manner) the exponential cutoff should be exact, provided there are no deeper non-spherical minima of the action, which appears rather unlikely. Besides, we have shown that this regime could be recovered from a simple spherical model. This is due to the spherical symmetry of the problem we investigate, which implies that in the limit of rare events the pdf is governed by rare almost spherical density fluctuations. 

Moreover, we have shown that our results agree with perturbative calculations over the range $-3<n<-1$ where the latter yield finite results (up to the first subleading term). Besides, since our method is non-perturbative it also applies to $-1 \leq n <1$ where shell-crossing comes into play (which leads to the break-up of perturbative approaches). This makes the derivation of the exact low-density tail more difficult but we obtained an approximate result which should provide the exact exponents which govern this low-density falloff. Note that in this regime our analysis shows that the spherical model only yields an approximate result for the low-density tail. Moreover, it clearly shows that perturbative results should be viewed with caution since leading-order terms may turn to be wrong (even when they are finite). This can be understood from the fact that perturbative expansions diverge (in fact, beyond a finite order one encounters divergent quantities, see \cite{paper5}). Next, we have pointed out that our results show that this low-density tail cannot be obtained from the stable-clustering ansatz which is often used to describe the highly non-linear regime ($\sigma \gg 1$). However, this does not imply that the latter model is wrong. It simply means that it is a zeroth-order approximation which does not capture the behaviour of these rare expanding voids. Finally, we have noticed that the low-density tail we described in this work is still out of reach of current numerical simulations. Therefore, because of the finite number of particles available in these numerical works, they cannot probe the actual low-density cutoff of the underlying continuous matter distribution yet.

Second, we have turned to the high density tail of the pdf. This case is much more difficult since shell-crossing now plays a key role. We have shown that a naive approach based on the exact spherical dynamics (which implies radial trajectories) fails. Indeed, a strong radial-orbit instability implies that the radial collapse solution is actually irrelevant. In particular, we have shown that collapsed halos see their velocity distribution become roughly isotropic over one dynamical time, whatever small the initial deviations from spherical symmetry. This leads to a very efficient virialization process, similar to ``violent relaxation''. Starting from an initial state which may be very far from thermodynamical equilibrium (in the sense that the transverse velocity dispersion may be very small) the halo relaxes over one dynamical time to a new equilibrium state where the velocity distribution is roughly isotropic. Moreover, we have shown that this process stabilizes the density profile. Thus, contrary to the cases of radial collapse or gaseous dynamics, the transverse velocity dispersion stabilizes new infalling shells to a finite radius so that the almost spherical halo obeys stable-clustering. 

Then, using these results we have derived the high-density tail of the pdf $\cP(\rho_R)$. The exponential cutoff can again be recovered from a simple spherical model (apart for a possible modification of the power-law prefactor). Moreover, this is consistent with the stable-clustering ansatz in the highly non-linear regime. We have also shown that these results are consistent with numerical simulations. Besides, it implies that the exponential cutoff of the standard Press-Schechter (PS) mass function is correct, although the prefactor and the characteristic density contrast $\dc$ may be modified. However, there is no reason to multiply the PS prescription by an ad-hoc factor 2 since it should not apply to low-mass halos. Similarly, our results only give the very high-density tail of the pdf (i.e. $\rho_R \gg \xib$).

In-between these extreme low-density and high-density tails of the pdf $\cP(\rho_R)$ there appears in the non-linear regime ($\sigma \ga 1$) an intermediate region which cannot be described by our steepest-descent approach. This range disappears in the quasi-linear regime ($\sigma \rightarrow 0$) as any finite density contrast becomes a rare event in this limit. This is why our approach can fully describe the quasi-linear regime as detailed in \cite{paper2} (see also \cite{paper3} for non-Gaussian initial conditions), in a way which is fully consistent with the study developed here. By contrast, in the non-linear regime this intermediate range corresponds to ``moderate'' density fluctuations whose dynamics is strongly influenced by their neighbours (e.g., tidal effects, mergings). Moreover, we argued that this regime might extend towards larger densities (in physical coordinates) as time goes on and the sensitivity to deviations from spherical symmetry grows. This would imply that the range of validity of the high-density tail we derived in this article would be repelled to increasingly large densities as one probes deeper into the highly non-linear regime. An understanding of this regime requires new non-perturbative tools which can handle the intricate non-local dynamics of these typical regions through strong tidal effects and merging processes. It is not clear at present whether this regime can be described by the stable-clustering ansatz, although numerical simulations suggest it should provide at least a good zeroth-order approximation.

Finally, we note that although we considered the case of a critical-density universe in this paper, it is straightforward to extend our results to arbitrary cosmological parameters $(\Om,\Ol)$. One simply needs to use the relevant function $\cFr(\dL)$ which describes the spherical dynamics (taking into account virialization). This does not introduce any new feature. Moreover, although we considered power-law linear power-spectra $P(k) \propto k^n$ our results can easily be extended to other smooth power-spectra like the usual CDM model. This merely modifies the characteristic function $\cGr(\tau)$ defined from the spherical dynamics function $\cFr(\dL)$, through the factor $\sigma(R_L)/\sigma(R)$ which is no longer a power of the overdensity. However, for practical purposes it should be sufficient to use the expressions derived here for power-law power-spectra, substituting for the index $n$ the local slope of $P(k)$ over the range of wavenumbers one is studying, as done for instance in \cite{Peac1} to obtain the non-linear two-point correlation function from its linear counterpart. Besides, it is clear that our results can also be extended to non-Gaussian initial conditions, like the models we studied in \cite{paper3}. In all cases, the simplest way to obtain the relevant expressions is to use the simple spherical model recalled in this article coupled with the pdf of the linear density field, in a fashion similar to the Press-Schechter prescription. This shows no difficulties. The only technical point is to derive the relevant linear pdf, but this can be done in a rigorous manner following the results detailed in \cite{paper3}, or through an appropriate application of the steepest-descent approach developed in \cite{paper2} to specific models.

\appendix

\section{Radial-orbit instability}
\label{Radial-orbit instability}

In this appendix, we investigate the stability of a spherically symmetric halo where all particles follow radial trajectories. Such a system arises from the collapse of an initial density fluctuation which obeys exact spherical symmetry for the linear growing mode. As expected, we shall find below that such a system exhibits a very strong radial-orbit instability with a growing rate $\om$ which goes to infinity as the trajectories become closer to radial. Thus, for our purposes we can restrict ourselves to small time intervals $\Delta t$ which are much smaller than the Hubble time $t_H$. Then, we can neglect the growth of the halo as new outer shells collapse on the time-scale $t_H$. Therefore, we study here the stability of a stationary spherical halo. If all orbits were exactly radial the equilibrium distribution function $f_0(\br,\bv)$ would be of the form $g(E) \delta_D(L^2)$, where $E$ is the energy of the particle and $L=|{\bf L}|$ the magnitude of the angular momentum. However, since perturbation theory is singular around such a state (as we shall see below) we add a small angular momentum to the particles. Thus, we write:
\beq
f_0(\br,\bv) = g(E) h_{\mu}(L)
\label{f01}
\eeq
with:
\beq
\int\d L^2 \; h_{\mu}(L) = \int\d L^2 \; \delta_D(L^2) =1
\eeq
and $h_{\mu}(L)$ is peaked for very small values of $L$ of order $\mu$. For instance, we write:
\beq
h_{\mu}(L) = \frac{1}{\mu^2} \; h\left(\frac{L}{\mu}\right) \;\; \mbox{with} \;\; \int_0^{\infty}\d x^2 \; h(x) = 1 ,
\label{h1}
\eeq
and we investigate the limit $\mu \rightarrow 0^+$. We use the spherical coordinates $(r,\theta,\phi;v_r,\vp,\alpha)$ with:
\beq
\vp = \sqrt{v_{\theta}^2 + v_{\phi}^2} , \; v_{\theta} = \vp \cos \alpha , \;  v_{\phi} = \vp \sin \alpha .
\label{alpha1}
\eeq
The angle $\alpha$ spans the range $0\leq \alpha<2\pi$. We note $\Phi_0(r)$ the equilibrium gravitational potential and $(f_1,\Phi_1)$ the perturbed distribution and potential at linear order. The linearized collisionless Boltzmann equation reads:
\beq
\left( \frac{\pl}{\pl t} + \bv . \nabla - \nabla \Phi_0 . \frac{\pl}{\pl \bv} \right) f_1 = \nabla \Phi_1 . \frac{\pl f_0}{\pl \bv}
\label{Bol1}
\eeq
In spherical coordinates the integration of eq.(\ref{Bol1}) yields (\cite{Frid1}):
\beq
f_1 = \int_{-\infty}^t \d t' \; \left( \frac{\pl \Phi_1}{\pl r} \frac{\pl f_0}{\pl v_r} + \frac{1}{r} (\hat{R} \Phi_1) \frac{\pl f_0}{\pl \vp} \right)
\label{f1}
\eeq
where the integration is taken along the non-perturbed trajectory and we introduced the linear operator $\hat{R}$ defined by:
\beq
\hat{R} \equiv \cos \alpha \frac{\pl}{\pl \theta} + \frac{\sin\alpha}{\sin\theta} \frac{\pl}{\pl \phi} - \sin\alpha \cot\theta \frac{\pl}{\pl \alpha} .
\label{R1}
\eeq
In eq.(\ref{f1}) we used the fact that $f_0(E,L)$ does not depend on $\alpha$. Of course, $\Phi_1(r,\theta,\phi,t)$ does not depend on $\alpha$ either. Next, using:
\beq
E = \frac{1}{2}(v_r^2+\vp^2) + \Phi_0(r) , \;\; L=r \vp ,
\label{Ev1}
\eeq
as well as:
\beq
\frac{\d \Phi_1}{\d t} = \frac{\pl \Phi_1}{\pl t} + v_r \frac{\pl \Phi_1}{\pl r} + \frac{\vp}{r} \hat{R} \Phi_1
\eeq
along the particle trajectory, we can write eq.(\ref{f1}) as:
\beqa
f_1(t) & = & \frac{\pl f_0}{\pl E} \left[ \Phi_1(t) - \int_{-\infty}^t \d t' \; \frac{\pl \Phi_1}{\pl t} \right] +  \frac{\pl f_0}{\pl L} \int_{-\infty}^t \d t' \; \hat{R} \Phi_1 . \nonumber \\
\label{f2}
\eeqa
Thus, once we are given a perturbed gravitational potential $\Phi_1(\br,t)$ the eq.(\ref{f2}) yields the associated distribution function $f_1(\br,\bv,t)$. Next, to obtain the eigenmodes of the system we simply need to take into account the Poisson equation. Indeed, the potential $\Phi_1(\br,t)$ must obey the consistency requirement:
\beq
\Delta \Phi_1 = 4\pi \cG \rho_1
\label{Poisson1}
\eeq
where the perturbed density $\rho_1$ is obtained from the distribution $f_1(\br,\bv,t)$ derived in eq.(\ref{f2}). Thus, substituting eq.(\ref{f2}) into eq.(\ref{Poisson1}) yields a linear eigenvalue problem for the unknown function $\Phi_1(\br,t)$. This procedure gives the eigenmodes of the system we look for. 

The radial-orbit instability will be produced by the last term in eq.(\ref{f2}). Indeed, when the typical angular momentum $\mu$ of the particles becomes very small the derivative $\pl f_0/\pl L$ gives rise to a factor $1/\mu$ which diverges in the radial limit. This is similar to the usual Jeans instability: in the limit $\mu\rightarrow 0$ the velocity dispersion in the transverse directions vanishes. Now, we look for the eigenmodes of eq.(\ref{f2}). Because of spherical symmetry the angular part of the perturbation can be decomposed over the spherical harmonics $Y_l^m(\theta,\phi)$. However, in order to simplify the analysis we use the three-index functions $T^l_{m,n}(\phi,\theta,\alpha)$ which provide a representation of the rotation of Euler angles $(\phi,\theta,\alpha)$, see \cite{Frid1} and \cite{Vil1}. Thus, we look for a perturbed gravitational potential of the form:
\beq
\Phi_1(\br,t) = e^{\om t} \; \chi(r) \; T^l_{m,0}(\phi,\theta,\alpha)
\label{Phi11} .
\eeq
Note that $T^l_{m,n}(\phi,\theta,\alpha)$ does not depend on $\alpha$ for $n=0$. Moreover, the functions $T^l_{m,0}$ are closely related to the usual spherical harmonics since we have $Y_l^m(\theta,\phi) \propto e^{2im\phi}T^l_{m,0}(\phi,\theta,\alpha)$. In particular, they also form a complete orthogonal system of functions on the sphere. The reason why we use the functions $T^l_{m,n}$ is that they provide a very convenient basis to write the action of the operator $\hat{R}$. Indeed, let us define the operators $\hat{H}_+$ and $\hat{H}_-$ by:
\beq
\hat{H}_+ \equiv e^{-i \alpha} \left[ i \frac{\pl}{\pl \theta} - \frac{1}{\sin\theta} \frac{\pl}{\pl \phi} + \cot\theta \frac{\pl}{\pl \alpha} \right]
\label{Hp1}
\eeq
and:
\beq
\hat{H}_- \equiv e^{i \alpha} \left[ i \frac{\pl}{\pl \theta} + \frac{1}{\sin\theta} \frac{\pl}{\pl \phi} - \cot\theta \frac{\pl}{\pl \alpha} \right] .
\label{Hm1}
\eeq
Then, the action of these operators $\hat{H}_+$ and $\hat{H}_-$ is simply (see \cite{Vil1}):
\beq
\hat{H}_+ T^l_{m,n} = - \sqrt{(l-n)(l+n+1)} \; T^l_{m,n+1} 
\label{Hp2}
\eeq
and:
\beq
\hat{H}_- T^l_{m,n} = - \sqrt{(l+n)(l-n+1)} \; T^l_{m,n-1} .
\label{Hm2}
\eeq
Next, the comparison of eq.(\ref{R1}) with eq.(\ref{Hp1}) and eq.(\ref{Hm1}) yields the relation:
\beq
\hat{R} = \frac{1}{2 i} \left( \hat{H}_+ + \hat{H}_- \right) .
\label{RH1}
\eeq
This gives the action of the operator $\hat{R}$ on the functions $T^l_{m,n}$ through eq.(\ref{Hp2}) and eq.(\ref{Hm2}). In particular, we obtain:
\beq
\hat{R} T^l_{m,0} = - \; \frac{\sqrt{l(l+1)}}{2 i} \left( T^l_{m,1} + T^l_{m,-1} \right) .
\eeq
Thus, eq.(\ref{f2}) reads:
\beqa
\lefteqn{ f_1(t) = \frac{\pl f_0}{\pl E} \left[  e^{\om t} \chi T^l_{m,0} - \int_{-\infty}^t \d t' \; \om e^{\om t'} \chi T^l_{m,0} \right] } \nonumber \\ & & - \frac{\sqrt{l(l+1)}}{2 i} \frac{\pl f_0}{\pl L} \int_{-\infty}^t \d t' \; e^{\om t'} \chi \left( T^l_{m,1} + T^l_{m,-1} \right) .
\label{f3}
\eeqa
We look for very large growth rates ($\om \rightarrow +\infty$) hence we expand eq.(\ref{f3}) over powers of $1/\om$. To do so, we use the relation (obtained from successive integrations by parts):
\beq
\int_{-\infty}^t \d t' \; e^{\om t'} f(t') = \frac{e^{\om t}}{\om} \left[ f(t) - \frac{f'(t)}{\om} + \frac{f''(t)}{\om^2} - ... \right]
\label{exp1}
\eeq
for arbitrary functions $f(t)$ such that the integral converges. Thus, up to second-order over $1/\om$ we obtain:
\beqa
\lefteqn{ f_1(t) = e^{\om t} \frac{\pl f_0}{\pl E} \left( \frac{1}{\om} \frac{\pl}{\pl t} - \frac{1}{\om^2} \frac{\pl^2}{\pl t^2} \right) (\chi T^l_{m,0}) } \nonumber \\ & & - \frac{\sqrt{l(l+1)}}{2 i} \frac{e^{\om t}}{\om} \frac{\pl f_0}{\pl L} \left( 1 - \frac{1}{\om} \frac{\pl}{\pl t} \right) ( \chi [T^l_{m,1} + T^l_{m,-1}] ) \nonumber \\
\label{f4}
\eeqa
The perturbed density $\rho_1(\br)$ is related to the distribution function $f_1$ by:
\beq
\rho_1(\br) = \int_{-\infty}^{\infty} \d v_r \int_0^{\infty} \frac{\d L^2}{2 r^2} \int_0^{2\pi} \d\alpha \; f_1(\br,\bv) .
\label{rho1}
\eeq
Besides, the functions $T^l_{m,n}(\phi,\theta,\alpha)$ are of the form:
\beq
T^l_{m,n}(\phi,\theta,\alpha) = e^{-i(m\phi+n\alpha)} \; P^l_{m,n}(\cos\theta)
\label{TP1}
\eeq
where the functions $P^l_{m,n}$ are closely related to the usual associated Legendre polynomials (\cite{Vil1}). Therefore, only the terms $T^l_{m,0}$ (i.e. $n=0$) contribute to the density $\rho_1$ after integration over the angle $\alpha$. This is the reason why we needed to go up to second-order over $1/\om$ in eq.(\ref{f4}) since the first-order term which is proportional to $\pl f_0/\pl L$ does not contain any factor $T^l_{m,0}$. 

Now, we must examine under which conditions the density $\rho_1$ is indeed dominated by the last term written in the expression (\ref{f4}). To do so, we must first recall the properties of the particle trajectories in a static spherical gravitational potential $\Phi_0(r)$. As is well-known, the orbit of a given particle is actually restricted to a plane which contains the origin (i.e. the center of the halo at $\br=0$) and which is orthogonal to the angular momentum $\bL$. Besides, the motion within this orbital plane can be described as a radial oscillation of frequency $\Omr$ coupled with an angular oscillation of frequency $\Omt$, with:
\beq
\frac{\pi}{\Omr} = \int_{r_{\rm min}}^{r_{\rm max}} \frac{\d r}{\sqrt{2(E-\Phi_0(r))-L^2/r^2}} ,
\label{Omr1}
\eeq
and:
\beq
\frac{\pi \Omt}{\Omr} = \int_{r_{\rm min}}^{r_{\rm max}} \frac{L \; \d r}{r^2\sqrt{2(E-\Phi_0(r))-L^2/r^2}} .
\label{Omt1}
\eeq
Here $r_{\rm min}$ and $r_{\rm max}$ are the minimum and maximum radii on the orbit. For extended halos we usually have $1/2 \leq \Omt/\Omr \leq 1$ since for a point mass (i.e. a Keplerian gravitational potential) we get $\Omt=\Omr$ while for a constant density (i.e. the harmonic oscillator) we have $\Omt=\Om/2$ (see \cite{Bin1}). Besides, for almost radial orbits (i.e. $L \rightarrow 0$) we usually have the resonance $\Omt=\Omr/2$ (the trajectory becomes almost symmetric with respect to the origin). For instance, we can check that this is the case for the density profile $\rho_0(r) \propto r^{-2}$ which would arise from eq.(\ref{rhor2}). This resonance leads to the usual radial-orbit instability investigated for instance in \cite{Pal1} and \cite{Pol1} (see also references therein). However, the instability we study here is rather different as it does not rely on this resonance. Indeed, these works considered the secular instability of nearly radial orbits within a halo where most particles have a significant angular momentum. In other words, the mass of the ``active'' particles (i.e. those with almost radial orbits which drive the instability) is small compared with the mass of the ``passive'' halo which determines the potential $\Phi_0$ (i.e. the particles with significant angular momentum which are stable). In such a case, the radial-orbit instability only involves a small fraction of the matter and it exhibits a slow growth rate $\om$, that is $\om \ll \Omo$ where $\Omo$ is the typical frequency of the system (i.e. $\Omo \sim \sqrt{\cG \rho_c}$ while the typical dynamical time is $t_D \sim 1/\sqrt{\cG \rho_c}$, where $\rho_c$ is the average density of the halo). Indeed, the instability is driven by the resonance $\Omt=\Omr/2$ so that the perturbation slowly increases through small cumulative effects over many orbital periods. By contrast, the system we investigate here is a halo where all particles are ``active'' and follow nearly radial orbits. Of course, this leads to a much more violent instability. In particular, we shall see below that we now obtain a very large growth rate $\om \gg \Omo$. This implies that the perturbation exhibits a strong growth in less than an orbital period. Therefore, the instability is not due to a resonance. In this sense, it is somewhat simpler and closer to the usual Jeans instability: it merely expresses the fact that in the absence of angular momentum there is no transverse velocity dispersion to stabilize the system against non-spherical perturbations.

From the properties of the nearly radial orbits we can obtain the magnitude of the various terms in eq.(\ref{f4}). Firstly, each derivative $\pl/\pl t$ (along the particle trajectory) yields a factor $\Omo$, where we note $\Omo$ the typical frequency of the system:
\beq
\Omo=\sqrt{\cG \rho_c} .
\label{Om01}
\eeq
Indeed, we noticed above that for radial orbits $\Omt=\Omr/2$ so that both frequencies are of the same order and close to $\Omo$. Therefore, in order to use the expansion (\ref{exp1}) we must have $\Omo/\om \ll 1$. Secondly, the second-order term in $\pl f_0/\pl L$ dominates over the first-order term in $\pl f_0/\pl E$ if $E_0 \gg \om \mu$. Here we note $E_0$ the typical energy of the particles which is of order $E_0 \sim (R \Omo)^2$, where $R$ is the radius of the halo. These two constraints yield:
\beq
\Omo \ll \om \ll \frac{L_0}{\mu} \; \Omo \;\;\; \mbox{with} \;\;\; L_0 = R^2 \Omo .
\label{limom1}
\eeq
This also implies:
\beq
\mu \ll L_0 .
\label{mu1}
\eeq
Note that $L_0$ is the typical angular momentum of generic orbits with $\vp \sim v_r$. Thus, we see at once from eq.(\ref{mu1}) that the radial-orbit instability we shall obtain below will stop when the particles exhibit significant transverse velocities. Therefore, the main effect of this process is to isotropize the velocity dispersion. In this respect, it is similar to violent relaxation: starting from a distribution function $f_0$ which is very far from thermodynamical equilibrium (since the transverse velocity dispersion $\sigma_{\perp}$ is zero) one obtains in a very short time (an orbital period !) a relaxed distribution where $\sigma_{\perp}$ is of the order of the radial velocity dispersion $\sigma_r$.

In the limit of nearly radial orbits ($\mu \ll L_0$) we can look for eigenmodes which obey the constraint (\ref{limom1}). Then, the density $\rho_1$ is dominated by the last term in eq.(\ref{f4}) which yields:
\beqa
\rho_1 = \frac{\sqrt{l(l+1)}}{2 i} \frac{e^{\om t}}{\om^2}  \int \frac{\d v_r \d L^2 \d\alpha}{2 r^2} \; \frac{\pl f_0}{\pl L} \chi \frac{\pl}{\pl t} (T^l_{m,1} + T^l_{m,-1}) . \nonumber \\
\label{rho2}
\eeqa
Thus, we now need the time derivative $\pl T^l_{m,n}/\pl t$. From the analysis of the trajectory in the orbital plane one can show that:
\beq
\frac{\pl}{\pl t} T^l_{m,n} = \Omt \; \hat{R} \; T^l_{m,n}
\label{dt1}
\eeq
where the operator $\hat{R}$ was defined in eq.(\ref{R1}) and $\Omt$ is the angular frequency of the orbit. Besides, using eq.(\ref{Hp2}) through eq.(\ref{RH1}) we have:
\beq
\int \d\alpha \; \hat{R} \; (T^l_{m,1} + T^l_{m,-1}) = - \frac{\sqrt{l(l+1)}}{i} \int \d\alpha \; T^l_{m,0}
\eeq
where we only need to keep the term $n=0$ since we integrate over $\alpha$. This yields:
\beq
\rho_1 = l (l+1) \pi \; \frac{e^{\om t}}{\om^2} \; \chi \; T^l_{m,0} \int \frac{\d v_r \d L^2}{2 r^2} \; \frac{\pl f_0}{\pl L} \; \Omt
\label{rho3} .
\eeq
As stated above, we can check in eq.(\ref{rho3}) that the spherical harmonics separate. Indeed, starting from a potential $\Phi_1$ of the form (\ref{Phi11}) we obtain a density $\rho_1$ which is proportional to the same harmonic $T^l_{m,0}$. Next, substituting the distribution function $f_0$ written in eq.(\ref{f01}) with eq.(\ref{h1}) we obtain:
\beqa
\rho_1 & = & - 2 \pi N_0 \frac{l(l+1)}{r^2\mu } \; \frac{e^{\om t}}{\om^2} \; \chi \; T^l_{m,0} \nonumber \\ & & \times  \int_{\Phi_0(r)}^{\Phi_0(R)} \frac{\d E}{\sqrt{2(E-\Phi_0(r))}} \; \Omt \; g(E)
\label{rho4}
\eeqa
where $N_0$ is a positive number of order unity given by:
\beq
N_0 = \int_0^{\infty} \d x \; h(x) , \;\; N_0 > 0 .
\label{N0}
\eeq
In eq.(\ref{rho4}) we used the fact that $h(L)$ is strongly peaked for small values of $L$ (of order of $\mu$ which obeys eq.(\ref{mu1})) so that $\Omt$ and $|v_r|$ in eq.(\ref{rho3}) are taken as functions of $E$ with $L=0$ (i.e. along radial orbits). Finally, we obtain the eigenmodes through the Poisson equation (\ref{Poisson1}). This yields:
\beqa
\lefteqn{ \frac{\d}{\d r} \left( r^2 \frac{\d \chi}{\d r} \right) - l(l+1) \chi = - 8 \pi^2 \cG N_0 \; \frac{l(l+1)}{\om^2 \mu} \; \chi } \nonumber \\ & & \times \int_{\Phi_0(r)}^{\Phi_0(R)} \frac{\d E}{\sqrt{2(E-\Phi_0(r))}} \; \Omt \; g(E)
\label{chi1}
\eeqa
which is a linear eigenvalue problem for the function of one variable $\chi(r)$. Thus, let us define the dimensionless linear operators $\hat{L_1}$ and $\hat{L_2}$ by:
\beq
(\hat{L_1} . \chi)(r) \equiv \frac{\d}{\d r} \left( r^2 \frac{\d \chi}{\d r} \right) - l(l+1) \chi(r)
\label{L1}
\eeq
and:
\beq
(\hat{L_2} . \chi)(r) \equiv - l(l+1) \; \zeta(r) \; \chi(r)
\label{L2}
\eeq
where we introduced the dimensionless function $\zeta(r)$ given by:
\beq
\zeta(r) = \frac{8 \pi^2 \cG N_0}{\Omo^2 \; L_0} \int_{\Phi_0(r)}^{\Phi_0(R)} \frac{\d E}{\sqrt{2(E-\Phi_0(r))}} \; \Omt \; g(E)
\label{zeta1}
\eeq
for $r<R$ and $\zeta(r)=0$ for $r>R$. Then, the linear equation (\ref{chi1}) reads:
\beq
\hat{L_1} . \chi = \left( \frac{\Omo}{\om} \right)^2 \; \left( \frac{L_0}{\mu} \right) \; \hat{L_2} . \chi
\label{chi2}
\eeq
Next, it is easy to see from eq.(\ref{L1}) and eq.(\ref{L2}) that for $l\geq 1$ both linear operators $\hat{L_1}$ and $\hat{L_2}$ are self-adjoint and negative definite over all functions $\chi(r)$ which are not identically zero over $0<r<R$, where we defined the scalar product:
\beq
\lag \chi_1 | \hat{L_i} | \chi_2 \rag = \int_0^{\infty} \d r \; \chi_1(r) (\hat{L_i} . \chi_2)(r)
\eeq
for any real functions $\chi_1$ and $\chi_2$. Indeed, the function $\zeta(r)$ defined in eq.(\ref{zeta1}) obeys $\zeta(r) >0$ for $r<R$ and $\zeta(r)=0$ for $r>R$. Therefore, if the function $\chi(r)$ is a solution of the eigenvalue problem (\ref{chi2}) we see that its associated growth rate $\om$ obeys $\om^2>0$, by taking the scalar product $\lag \chi | \hat{L_1} | \chi \rag$ in eq.(\ref{chi2}). Indeed, if $\chi(r)$ is zero over $0<r<R$, then we have $\hat{L_1}.\chi=0$ over $[0,+\infty[$ which implies $\chi=0$ (using the boundary conditions $\chi(r) \rightarrow 0$ for $r \rightarrow \infty$ and $\chi(r=0)$ finite). Thus, we get a real growth rate $\om >0$ which implies that the system is unstable for non-spherical perturbations (i.e. $l\geq 1$). Finally, the eigenmodes may be obtained from the more usual eigenvalue problem:
\beq
\hat{L} . \chi = \left( \frac{\Omo}{\om} \right)^2 \; \left( \frac{L_0}{\mu} \right) \; \chi
\label{chi3}
\eeq
where we introduced the linear operator:
\beq
\hat{L} \equiv \hat{L_2}^{-1} . \hat{L_1} = \frac{1}{\zeta(r)} \; \left[ 1 - \frac{1}{l(l+1)} \frac{\d}{\d r} \left( r^2 \frac{\d}{\d r} \right) \right]
\label{L1L2} .
\eeq
Thus, one first solves eq.(\ref{chi3}) over $0<r<R$, which yields a continuous spectrum of functions $\chi(r)$, and next the match at $R$ with the standard decaying solution of $\hat{L_1}.\chi=0$ defined over $r>R$ gives rise to a discrete spectrum. Let us note $\lambda_k$ the discrete eigenvalues of the operator $\hat{L}$ obtained in this way. As shown above, they are positive dimensionless numbers which give the associated growth rates through:
\beq
\om_k = \Omo \; \sqrt{\frac{L_0}{\lambda_k \mu}} .
\label{omk1}
\eeq
Thus, we see that the growth rates are of order $\om \sim \Omo \sqrt{L_0/\mu}$. Therefore, they satisfy the condition (\ref{limom1}) when $\mu \ll L_0$, which validates the calculation described above. Note that perturbation theory is singular for a system with exactly radial orbits since the growth rates $\om$ diverge in the limit $\mu \rightarrow 0$. This is why we introduced a small angular momentum $\mu$. On the other hand, we stress that the analysis described above is quite general. Thus, it does not rely on the shape of the equilibrium density profile $\rho_0$ nor on the distribution function $f_0$. Indeed, it is only based on the conditions (\ref{limom1}) and (\ref{mu1}). Therefore, the system remains unstable until the constraint (\ref{mu1}) is violated. This means that the halo quickly reaches an equilibrium state where the radial and transverse velocities are of the same order.


\begin{thebibliography}{}

\bibitem[Balian \& Schaeffer (1989)]{Bal1}
Balian, R., Schaeffer, R., 1989, A\&A 220, 1

\bibitem[Bernardeau (1994)]{Ber1}
Bernardeau, F., 1994, ApJ 427, 51

\bibitem[Bertschinger (1985)]{Bert1}
Bertschinger, E., 1985, ApJS 58, 39

\bibitem[Binney \& Tremaine (1987)]{Bin1}
Binney, J., Tremaine, S., 1987, Galactic dynamics, Princeton University Press 

\bibitem[Cole \& Lacey (1996)]{Cole1}
Cole, S., Lacey, C., 1996, MNRAS 281, 716

\bibitem[Colombi et al. (1996)]{Col1}
Colombi, S., Bouchet, F.R., Hernquist, L., 1996, ApJ 465, 14

\bibitem[Fillmore \& Goldreich (1984)]{Fil1}
Fillmore, J.A., Goldreich, P., 1984, ApJ 281, 1 

\bibitem[Fridman \& Polyachenko (1984)]{Frid1}
Fridman, A.M., Polyachenko, V.L., 1984, Physics of gravitating systems, Springer-Verlag

\bibitem[Governato et al. (1999)]{Gov1}
Governato, F., Babul, A., Quinn, T., Baugh, C.M., Katz, N., Lake, G., 1999, MNRAS 307, 949

\bibitem[Gradshteyn \& Ryzhik (1965)]{Grad1}
Gradshteyn, I.S., Ryzhik, I.M., 1965, Table of integrals, series and products, fourth edition, Academic Press

\bibitem[Hamilton et al. (1991)]{Ham1}
Hamilton, A.J.S., Kumar, P., Lu, E., Matthews, A., 1991, ApJ 374, L1

\bibitem[Jain et al. (1995)]{Jain1}
Jain, B., Mo, H.J., White, S.D.M., 1995, MNRAS 276, L25

\bibitem[Kandrup \& Sygnet (1985)]{Kan2}
Kandrup, H.E., Sygnet, J.F., 1985, ApJ 298, 27

\bibitem[Kandrup et al. (1993)]{Kan1}
Kandrup, H.E., Mahon, M.E., Smith, H., 1993, A\&A 271, 440

\bibitem[Palmer \& Papaloizou (1987)]{Pal1}
Palmer, P.L., Papaloizou, J., 1987, MNRAS 224, 1043

\bibitem[Peacock \& Dodds (1996)]{Peac1}
Peacock, J.A., Dodds, S.J., 1996, MNRAS 280, L19

\bibitem[Peebles (1980)]{Peeb1}
Peebles, P.J.E., 1980, The large scale structure of the universe,
Princeton University Press

\bibitem[Polyachenko (1992)]{Pol1}
Polyachenko, V.L., 1992, Sov. Phys. JETP 74 (5), 755

\bibitem[Press \& Schechter (1974)]{PS}
Press, W.H., Schechter, P., 1974, ApJ 187, 425

\bibitem[Scoccimarro \& Frieman (1996)]{Scoc1}
Scoccimarro, R., Frieman, J.A., 1996, ApJ 473, 620

\bibitem[Teyssier et al. (1997)]{Tey1}
Teyssier, R., Chieze, J.-P., Alimi, J.-M., 1997, ApJ 480, 36

\bibitem[Tormen et al. (1997)]{Tor1}
Tormen, G., Bouchet, F.R., White, S.D.M., 1997, MNRAS 286, 865

\bibitem[Valageas \& Schaeffer (1997)]{Val1}
Valageas, P., Schaeffer, R., 1997, A\&A 328, 435

\bibitem[Valageas (1998)]{Val2}
Valageas, P., 1998, A\&A 337, 655

\bibitem[Valageas et al. (2000)]{Lac1}
Valageas, P., Lacey, C., Schaeffer, R., 2000, MNRAS 311, 234

\bibitem[paper I]{paper1}
Valageas, P., 2001, A\&A 379, 8, Paper I

\bibitem[paper II]{paper2}
Valageas, P., 2001, accepted by A\&A, Paper II [astro-ph/0107126]

\bibitem[paper III]{paper3}
Valageas, P., 2001, accepted by A\&A, Paper III [astro-ph/0107196]

\bibitem[paper V]{paper5}
Valageas, P., 2001, accepted by A\&A, Paper V [astro-ph/0109408]

\bibitem[Vilenkin (1968)]{Vil1}
Vilenkin, N.J., 1968, Special functions and the theory of group representations, American mathematical society, Providence, Rhode Island

\end{thebibliography}
\end{document}